\documentclass{aa}  

\usepackage{graphicx}
\usepackage{xcolor}

\usepackage{txfonts}

\newcommand{\ks}{{$\rm K_S$}\ }
\newcommand{\ksnosp}{{$\rm K_S$}}
\newcommand{\jks}{{$\rm J-K_S$}\ }
\newcommand{\jksnosp}{{$\rm J-K_S$}}
\newcommand{\ejks}{{$\rm E(J-K_S)$} }
\newcommand{\ejksnosp}{{$\rm E(J-K_S)$}}

\begin{document} 

\title{VVV catalog of ab-type RR Lyrae in the inner Galactic bulge}

   \author{M. Zoccali\inst{1,2}
          \and
          C. Quezada\inst{1,2}
          \and
          R. Contreras Ramos\inst{1,2}
          \and
          E. Valenti\inst{3,4}
          \and
          A. Valenzuela-Navarro\inst{1,2}
          \and
          J. Olivares Carvajal\inst{1,2}
          \and
          A. Rojas Arriagada\inst{2,5,6,7}
          \and
          J. H. Minniti\inst{2,8}
          \and
          F. Gran\inst{9}
          \and
          M. De Leo\inst{1,2}
          \fnmsep\thanks{Based on observations taken within the ESO VISTA Public Survey VVV, Program ID 179.B-2002.}
          }

\institute{Instituto de Astrof\'isica, Pontificia Universidad Cat\'olica de Chile, Av. Vicu\~na Mackenna 4860, 782-0436 Macul, Santiago, Chile
\email{mzoccali@uc.cl}
    \and
Millennium Institute of Astrophysics, Av. Vicu\~na Mackenna 4860, 82-0436 Macul, Santiago, Chile 
    \and
European Southern Observatory, Karl Schwarzschild-Strabe 2, 85748 Garching bei München, Germany
    \and
Excellence Cluster ORIGINS, Boltzmann\--Stra\ss e 2, D\--85748 Garching bei M\"{u}nchen, Germany
    \and
Departamento de F\'isica, Universidad de Santiago de Chile, Av. Victor Jara 3659, Santiago, Chile
    \and
N\'ucleo Milenio ERIS
    \and
Center for Interdisciplinary Research in Astrophysics and Space Exploration (CIRAS), Universidad de Santiago de Chile, Santiago, Chile
    \and
Department of Physics and Astronomy, Johns Hopkins University, Baltimore, MD 21218, USA
    \and
Universit\'e C\^ote d’Azur, Observatoire de la C\^ote d’Azur, CNRS, Laboratoire Lagrange, Nice, France
}

   \date{Received September 15, 1996; accepted March 16, 1997}

  \abstract
{Observational evidence has accumulated in the past years, showing that the Galactic bulge includes two populations, a metal poor and a metal rich one that, in addition to a different metallicity, show different alpha over iron abundances, spatial distribution, and kinematics. While the metal rich, barred component has been fairly well characterized, the metal poor, spheroidal component has been more elusive and harder to describe. RR Lyrae variables are clean tracers of the old bulge component, and they are, on average, more metal poor than red clump stars. 
}  
{In the present paper, we provide a new catalog of 16488 ab type RR Lyrae variables in the bulge region within |l|$\lesssim $10$^\circ$ and |b|$\lesssim $2.8$^\circ$, extracted from multi epoch PSF photometry performed on VISTA Variable in the V\'\i a L\'actea survey data. We used the catalog to constrain the shape of the old, metal poor, bulge stellar population.
}
{The identification of ab type RR Lyrae among a large sample of candidate variables of different types has been performed via a combination of a Random Forest classifier and visual inspection. We optimized this process in such a way to extract a clean catalog with high purity, although for this reason its completeness, close to the midplane, is lower compared to a few other near infrared catalogs covering the same region of the sky. }
{We use the present catalog to derive the shape of their distribution around the Galactic Center, resulting in an elongated spheroid with projected axis ratio b/a$\sim$0.7 and inclination angle $\phi$$\sim$20 degrees. We discuss how observational biases such as errors on the distances
and a non-uniform sampling in longitude, affect both the present measurements and previous ones, especially those based on red clump stars. Because the latter have not been taken into account before, we refrain from a quantitative comparison between these shape parameters and those derived for the main Galactic bar. Nonetheless, qualitatively, taking into account observational biases would lower the estimated ellipticity of the bar derived from RC stars, hence reducing the difference with the present results.}
{We publish a high purity RRab sample for future studies of the oldest Galactic bulge population, close to the midplane. We explore different choices for the period-luminosity-metallicity relation, highlighting how some of them introduce spurious trends of distances with either period or metallicity, or both.
We provide evidence that they trace a structure less elongated than the main bar, though we also highlight some biases of these kind of studied not discussed before. }


   \maketitle
%
\section{Introduction}

In the context of the Milky Way (MW) formation, the study of the Galactic bulge is especially relevant as it is the
first massive component to get into place. Yet, it is possibly the component whose formation we understand the least.
Until about one decade ago, the paradigm for the formation of bulges was built upon the dichotomy postulated by \cite{kk+04}, between the so-called {\it classical} bulges, formed by dissipationless merging of small external galactic fragments, and the bar-like {\it pseudo-}bulges, originating from disk instabilities. More recently, observations of high redshift (z$\sim$2) star forming disks showed the presence of dense clumps with high star formation rates (SFR) \citep{Guo+15, Guo+18, Dessauges+17, Cava+18, Huertas+20}. A big such clump is usually present at the center of the disks, obviously showing a massive, early, in situ star formation event. Thanks to these findings, the {\it classical/pseudo} bulge dichotomy has been abandoned in favor of scenarios where bulges form mostly in situ, but on a short (definitely non-secular) timescale \citep[e.g.,][]{fragkoudi+20}.

Recently, \citet{debattista+23} presented N-body + smoothed particle hydrodynamics simulations that develop high
SFR clumps, including a large one at the center. Some of the clumps formed in the disk eventually migrate towards the
center, where they rapidly ($<$1 Gyr) dissolve. \citet{debattista+23} demonstrated that such clumpy models are able to explain the bimodal distribution in [Fe/H], observed for bulge stars, together with the observed presence of two density peaks, separated by a trough, in the [$\alpha$/Fe] $vs$ [Fe/H] plane. Indeed, according to these models, stars formed
in clumps have relatively large [$\alpha$/Fe] ratios, while stars formed in the field, with lower SFRs, result
in comparatively lower [$\alpha$/Fe] ratios. While this represents an important step forward towards qualitatively 
reconciling the observations of high redshift disks with those of individual stars in the MW, several issues
remain to be understood.

For instance, one important point to be addressed is why the metal poor stars ([Fe/H]$<$0) in the bulge seem to be arranged in a rather axisymmetric component, while the metal rich stars ([Fe/H]$>$0), trace a well defined bar \citep[e.g.,][]{zoccali+17, lim+21}. Note that, long before it was demonstrated that the metal-poor stars traced a component that is more axisymmetric than the bar, a few independent studies had shown that they did not show the X-shape structure either \citep{ness+12, rojas-arriagada+14}.
In what follows, the axisymmetric component will be called "the spheroid", although its shape has been inferred only very qualitatively.

Models by \citet{diMatteo+16, debattista+17, fragkoudi+18} are able to explain the origin of both bulge components as deriving completely from the disk, where a hotter (e.g., thick) disk may naturally generate the axisymmetrical structure, while a colder (e.g., thin) disk would map into a more pronounced bar. This process has been dubbed "kinematical fractionation" by \citet{debattista+17}. A comprehensive model combining the disk fractionation needed to produce two different spatial structures, with the presence of clumps needed to produce the bimodality in the [$\alpha$/Fe] trends, is not yet available.

To further complicate things, \cite{queiroz+21} recently combined chemical abundances and radial velocities from APOGEE DR17 with proper motions from Gaia DR3 to derive orbits. They concluded that both the stars with bar-like orbits and those with spheroid-like orbits are uniformly scattered in the [Mg/Fe] versus [Fe/H] plane. This finding, if confirmed, would dismantle the scenario where stars in the spheroid are alpha-rich because they came from the thick disk, while stars in the bar are alpha-poor because they have a thin disk origin.

To progress on this topic, an accurate characterization of the spheroid is highly desirable. Currently, several works have used red clump (RC) stars to derive the shape parameters of the bar. All of them, however, used samples of RC stars that include both metal rich and metal poor stars in roughly equal proportion \citep{zoccali+18}. Therefore, we know that the derived bar parameters are not completely correct. In order to separate the two components, metallicities are needed, which are not available for the large samples. In other words, nobody has yet attempted to isolate stars belonging to the metal-poor component (the spheroid) and derive, e.g., the radial density profile or axial ratio of their parent population. A recent catalog providing photometric metallicities for 2.6 million RC stars, at latitudes $-$10$<$b$<$$-$4, has been published by \citet{johnson+22}. As discussed in \cite{lim+21}, these data qualitatively confirm the results by \citet{zoccali+17} about the spheroid (bar) shape traced by metal-poor (metal-rich) RC stars, but a full derivation of the structural parameters of each of the two components has not been made with these new data, possibly due to their limited latitude coverage.

RR Lyrae (RRL) variables are of great value in this context because they trace a pure old population \citep[$>$ 10 Gyr;][]{walker+89}, on average more metal-poor than most RC stars. As such, they might be clean tracers of the spheroid. In addition, thanks to a well defined relation between their periods and their absolute luminosity (PL), they provide much more precise distances than RC stars \citep[see, e.g.,][]{girardi16}
allowing us to trace the three-dimensional (3D) shape of their parent population with greater confidence. 

Results in the literature on this point are rather controversial. By analyzing the 3D distribution of RRL from the optical OGLE III survey, \cite{pietrukowicz+12} first concluded that they tend to follow the Galactic bar traced by RC stars, a result later confirmed in \cite{pietrukowicz+20} based on improved data from OGLE-IV. Meanwhile, \cite{dekany+13} combined the OGLE III RRL catalog with near infrared (IR) magnitude measurements from the Vista Variables in the V\'ia L\'actea (VVV) ESO Public Survey \citep{minniti10}, less sensitive to extinction, and found a more axisymmetric, mildly barred component. 
By analyzing an independent sample of $\sim$1000 RRL observed by VVV at latitude $-10.3<$b$<-8.0$, a region not explored by OGLE, \citet{gran+16} claimed the absence of a bar nor an X-shape. \citet{semczuk+21}, on the contrary, did find the signature of an X-shape in the OGLE-IV data. 
Recently, \citet{du+20} show that the shape parameters of the RRL parent population depend on their metallicity: metal rich RRL ([Fe/H]$>$$-$1) define a stronger
bar compared to metal-poor ones. Additionally, \citet{kunder+16}, found that RRL stars show different dynamical properties than metal-rich RC stars, more consistent with the Galactic bulge spheroid.

In the present paper, we provide a new catalog of (fundamental mode) ab type RRL variables (hereafter RRab) in the inner bulge region extracted from VVV PSF photometry (see Sec.~\ref{sec:data}). The classification of their light curves, described in Sec.~\ref{sec:RF}, has been optimized to obtain a pure sample, though with lower completeness compared to other recent near IR catalogs (see Sec.~\ref{sec:othercats}). We excluded first overtone, c-type, RRL because their lower amplitude and symmetric light curve make them easily confused with eclipsing binaries. Distances are derived in Sec.~\ref{sec:metdist}, and their spatial distribution is analyzed in Sec.~\ref{sec:gcenter}
and Sec.~\ref{sec:3D}. Conclusions are drawn in Sec.~\ref{sec:summary}.

\section{Observations and Photometry}  
\label{sec:data}

\begin{figure}
    \includegraphics[width=\hsize,trim = {1cm 1cm 3cm 2cm}, clip]{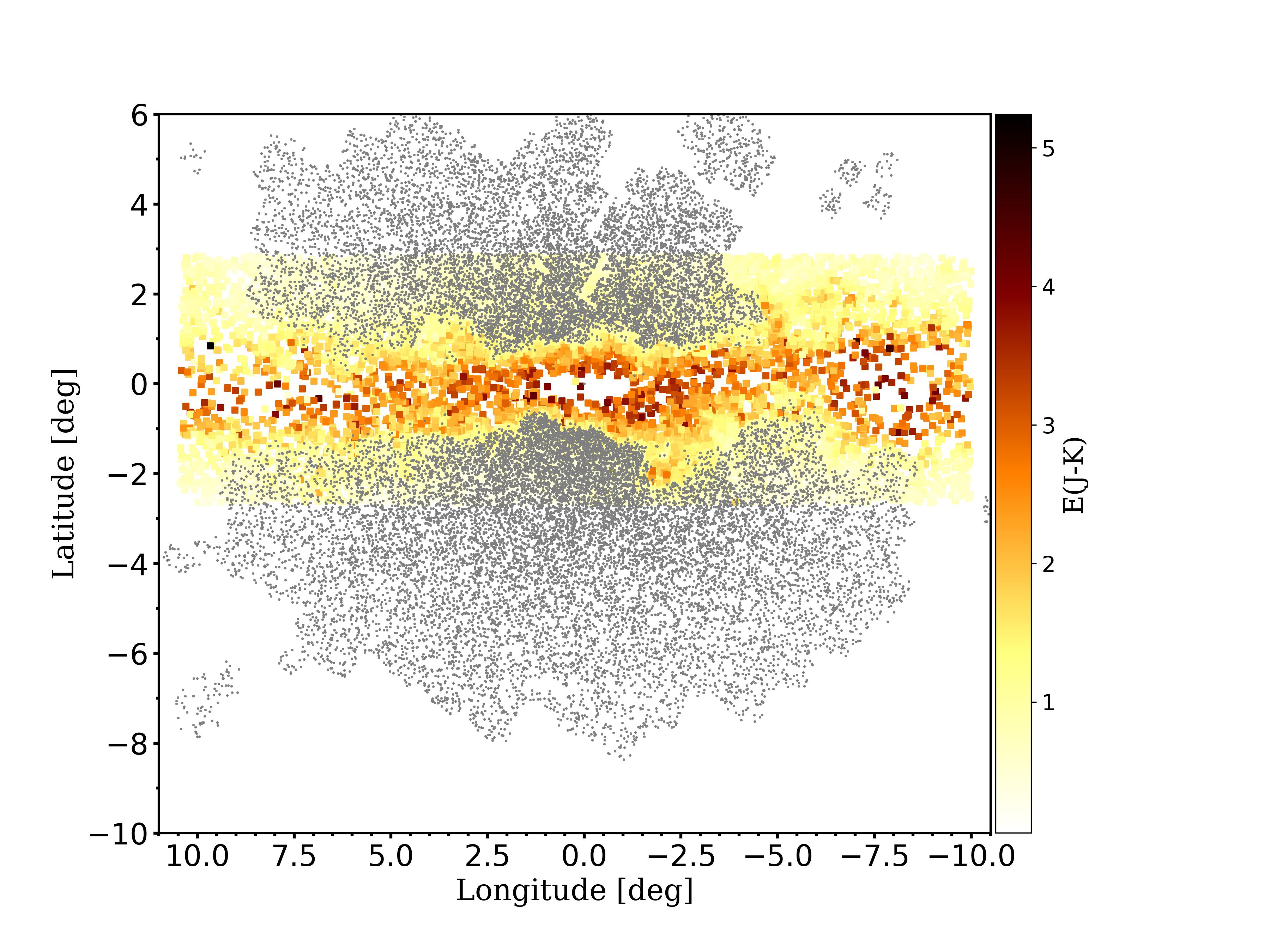}
    \caption{Galactic coordinates of the 16488 RRL in the catalog presented here, color-coded according to their extinction \ejks. Overlaid in grey are the coordinates of RRab variables in the OGLE IV catalog.}
    \label{fig:sky_area}
\end{figure}

The catalog of RRL presented here is based on PSF--fitting photometry performed on multi epoch near IR data from the VVV ESO Public Survey. The survey has observed an area of about 1700 deg$^{2}$ over the MW bulge and part of the southern disk since 2010. For the present study, we only analyzed the region within $-10^\circ$$\lesssim $l $\lesssim $10.5$^\circ$ and $-2.6^\circ$$\lesssim $b $\lesssim $2.8$^\circ$ (Fig.~\ref{fig:sky_area}), i.e., from tile b299 to b368. This is the region where previous optical surveys such as OGLE IV (Fig.~\ref{fig:sky_area}) or Gaia (Fig.~\ref{fig:gaia}),
are highly incomplete or absent, due to the high interstellar extinction close to the Galactic midplane. PSF photometry was performed on all the available epochs, i.e., 100 in \ks and less than a dozen in J and H. The time baseline spans 10 years, from 2010 to 2019. The typical seeing is close to 1 arcsec; the few images with seeing larger than 1.7 arcsec were discarded from the analysis. This limit 
has been chosen in order to reduce blends in the very crowded region examined here, at the same time excluding only a few images per field.

The multi epoch photometry provided us with both stellar variability and proper motions. A photometric pipeline was built, based on the DAOPHOT II/ALLSTAR package \citep{stetson87}, optimized for the measurement
of proper motions following the prescriptions by \cite{anderson06} and \cite{bellini14}. The exact recipe is described in \citet{contrerasramos17}. Here, we recall just the main steps. A quick photometry, with a very high threshold, is performed initially to detect only relatively bright stars. The latter are used to derive coordinate transformations among all the epochs, for each VIRCAM chip. A stacked image is then created using those transformations, and a deep photometry is performed on it, with a small threshold, in order to create the most complete stellar master list. This list of stars is then used as input for the PSF photometry in each individual epoch, allowing the code to refine the centroids of each star. Consequently, each given star has the same ID in all the epochs, but its position is allowed to vary with time, in order to measure the star's proper motion. To look for variable stars, all the instrumental magnitudes measured for a given star are normalized to the photometric system of an arbitrary reference epoch. Therefore, only this epoch must be calibrated to a standard photometric system.

\begin{figure}
	\includegraphics[width=\hsize]{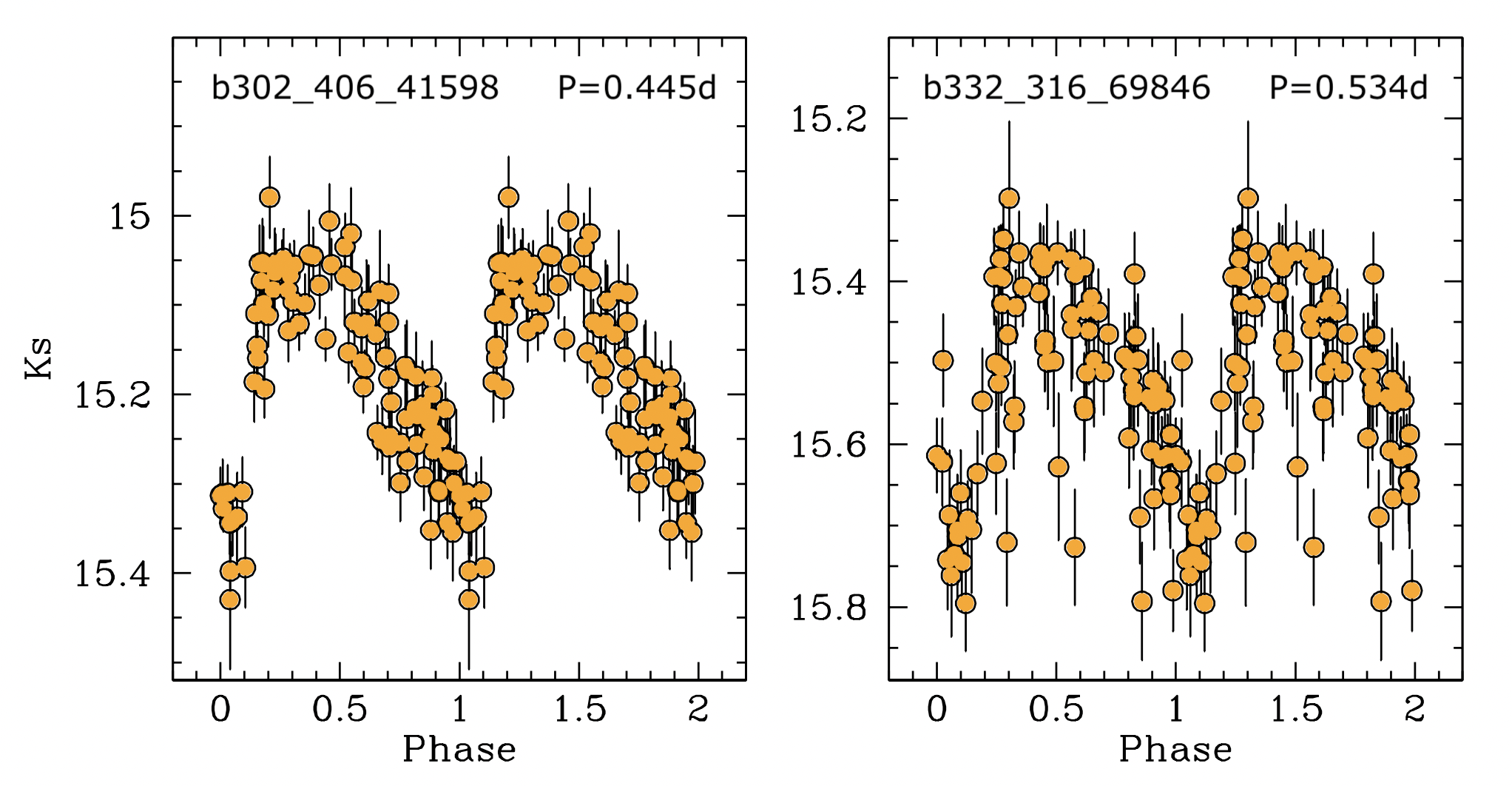}
    \caption{Phased, \ks-band light curves for two typical RRab variables in our sample.}    
    \label{fig:lc}
\end{figure}

In our case, we initially calibrated the catalogs to the VISTA photometric system \citep{gonzalez-fernandez18}, by means of a large number of stars in common with the public catalogs, which are based on aperture photometry, released by the Cambridge Astronomical Survey Unit \citep[CASU]{emerson04}.  Recently, however, \cite{hajdu+20} noted that the standard CASU calibration procedure has zero-point inaccuracies, especially in regions of high stellar density. We thus recalibrated the photometry following the general prescription described in their work, i.e., we matched our catalogs directly with the 2MASS ones, imposing rather conservative -and supervised- cuts in order to discard stars that might be blended in 2MASS. A complete description of this procedure can be found in \cite{nikzat22}. 

Example of the final photometry is presented in Fig.~\ref{fig:lc}, showing the light curves of two typical RRab in our sample. The left one is in tile b302, with approximate coordinates (l,b)=($-$5$^\circ$,$-$2$^\circ$), while the one on the right is tile b332, at coordinate (l,b)=($-2$$^\circ$, 0$^\circ$). The latter is more noisy due to the increasing crowding and extinction when moving close to the Galactic center (GC).



\section{Identification of RR Lyrae variables}
\label{sec:RF}

Our PSF photometry detected approximately $3\times10^8$ point sources in the 70 VVV tiles analyzed here. In order to select candidate variable stars with well-measured magnitudes, first, we discarded measurements with unusually large photometric errors (e\ks$>0.1$\,mag) and applied an iterative 8$\sigma$ clipping to their (non-phased) light curves to reject outliers. We further omitted candidate variable stars with less than  50 measurements (i.e., less than $\sim$$50\%$ of the available epochs), and those with total amplitude (max$-$min) smaller than 0.1 mag. The former criterion was chosen because VVV light curves are unevenly sampled, and, close to the midplane, they are relatively noisy (Fig.~\ref{fig:lc}). The latter criterion ensures that we keep all the RRab, but exclude variables with amplitudes too small compared with the VVV typical photometric errors \citep[c.f., Fig.~8 in][]{contrerasramos18}.

At this point, in order to select candidate intrinsic variables, we run the (slightly modified) reduced $\chi^{2}$ test \citep{carpenter01} for all the time series. This value was computed, for each star, from the individual magnitudes (\ksnosp) and the corresponding uncertainties (e\ksnosp), as follows:

\begin{equation} \label{eq-chi2}
    \chi_{\nu}^{2} =  \frac{\sum_{i=1}^N (K_{s_{i}} - \bigl \langle wK_{s} \bigr \rangle)}{(N-1) \sum_{i=1}^N  eK_{s_{i}}}  
\end{equation}
\\
where $w$\ks is the error-weighted arithmetic mean of the \ks magnitude, and N is the number of epochs. Sources with a $\chi_{\nu}^{2}$ value greater than 3 were considered candidate variable stars.

In the second step, we used the Analysis of Variance \citep[AoV;][]{schwarzenberg-czerny89} method to compute periods for all the variable candidates that passed the first selection. The AoV also provides a quality index (Q) for each phase-folded light curve, using the period with the highest probability. After visually inspecting a few dozen cases, we adopted Q=0.9 as our cutoff value to select candidate variables. Our goal is to provide a high purity RRL catalog with minimal contamination from different type of variables, even at the expenses of completeness. In order to do so, we decided to select only RRab variables, because the lower amplitude and more symmetric light curves of the c-type RRL (RRc) makes them more easily mistaken with other variables, particularly eclipsing variables. Therefore, we selected stars with 0.3$<$P$<$1 days in order to make sure we include all RRab, at the same time avoiding short-period Cepheids and minimizing contamination from long-period RRc variables. Note that the latter comprises $\lesssim $30$\%$ of all RRL \citep{martinez-vazquez+17,soszynski+19,stringer+21}.

After the selection cuts described above, we were left with a sample of approximately 10 million candidate variable stars. The next and final step in our selection procedure was to apply a machine-learning classifier to these candidates, which we describe in the next section.

\subsection{Random Forest} 

Identifying RRL variables is more challenging in the near IR, compared to the optical, because at longer wavelengths, they show smaller amplitudes and a more sinusoidal (i.e., featureless) light curve \citep[see, e.g., Fig.~2 in][]{bhardwaj+22}. For this reason, they are easily confused with other kinds of variables, particularly eclipsing binaries. Due to the huge number of candidates selected in the central region, we developed a simple machine learning, supervised classifier based on Random Forest \citep{RFcode} from the Python library sci-kit learn \citep{scikit-learn}. Random Forest classification requires three main sets of input data from the user: the training catalog, with different kinds of variables already identified; the science catalog, and a list of features (properties) quantitatively describing the brightness variation.  The code is requested to distinguish RRL from other variables in the science catalog, learning how the features behave in the training catalog, also called the training set. The code randomly builds a number of binary decision trees, where each object is classified according to a handful of features across each tree. The final classification is based on the average of the result of a large number (in our case, 20,000) of randomly built decision trees.

Random Forest is a relatively simple classification algorithm compared with more modern machine learning codes. However, since, in this case, we were interested in isolating only RRL variables, we built a very basic code that was required to discriminate only between RRL and "other variables", which turned out to work fine for our purposes.

\subsubsection{The training set} 
The training set consisted of more than 65,000 variable stars of different types, identified and characterized within the OGLE-IV survey \citep{ogleIV}. For all of them, we used the \ks band light curve from VVV, but the period was measured in OGLE. Specifically, the training set included $\sim$19000 RRab, $\sim$300 c-type RRL, $\sim$30000 non-contact and $\sim$16000 contact eclipsing variables, and $\sim$800 elipsoidal variables. They all had periods $<$2 days, according to OGLE. 

For the present work, we aimed at identifying only bona fide RRab (P=0.30$-$1.0 days), i.e., we excluded c-type (P=0.2$-$0.45 days) and d-type RRL. The latter two types, which are pulsating in the first overtone and in a mixed mode, respectively, are much less abundant than the former, fundamental mode pulsators \citep[e.g.,][]{smith95,catelansmith15}. Their light curves are sinusoidal with lower amplitude and can be easily confused with the more abundant short-period eclipsing binaries. For this reason, we decided to exclude them from the present study, and therefore the c-type RRL present in the training set are considered "other variables".

\subsubsection{The features}

\begin{table*}
\caption{Features used in the Random Forest classification.}
\begin{tabular}{l|l}
\hline\hline
\multicolumn{2}{c}{Group 1: Magnitude distribution features}   \\
\hline
Feature name & Description \\
\hline 
\small
Sigma           & The standard deviation of all the magnitudes about the mean K$_s$ band magnitude$^2$ \\
Skewness        & The skewness of the magnitude distribution$^2$ \\
Kurtosis        & The kurtosis of the magnitude distribution$^2$ \\
MV              & The ratio between the Sigma and the median magnitude in the K$_s$ band$^3$ \\
AD              & The Anderson-Darling test$^1$ assessing how similar the distribution is to a Gaussian \\
AC$_{\rm std}$  & The standard deviation of the convolution of the light curve with itself$^3$ \\
Slope           & The slope of a linear fit to the (non-phased) light curve$^2$ \\
MAD             & The Mean Absolute Deviation about the Median defined above$^2$ \\
RCS             & The ratio between the MAD and the Sigma defined above$^3$ \\
Amplitude (A)   & The difference between the magnitudes at the 95 and 5 percentile. A sort of robust amplitude$^2$ \\
mpr20           & The ratio between the magnitude at 60 and 40 percentile, and the amplitude defined above$^2$ \\
mpr35           & The ratio between the magnitude at 67 and 32 percentile, and the amplitude defined above$^2$ \\
mpr50           & The ratio between the magnitude at 75 and 25 percentile, and the amplitude defined above$^2$ \\
mpr65           & The ratio between the magnitude at 82 and 17 percentile, and the amplitude defined above$^2$ \\
mpr80           & The ratio between the magnitude at 90 and 10 percentile, and the amplitude defined above$^2$ \\
Beyond1Std      & Percentage of points beyond one st. dev. from the weighted mean \\
& The ratio between the magnitudes below over above 1 sigma from the mean (weighted by the photometric error)$^2$\\
Tm              & Ratio between the difference {\it minimum-mean}, over the total amplitude $^4$ \\

\hline
\multicolumn{2}{c}{Group 2: Shape of the light curve features}   \\
\hline
Period                  & Period \\
$A1A2_{ratio}$          & The ratio of the amplitudes when the light curve is phased with the double of the period (discards binaries with \\
 & different minima) \\
Rising Time             & Phase difference between the first minimum to the next maximum of the light curve \\
Falling Time            & Phase difference between the absolute maximum and the next minimum of the light curve \\
Rising/Falling time     & Ratio between the two quantities defined above \\
Phased Skewness         & The skewness of the phased light curve \\
\hline
\multicolumn{2}{c}{Group 3: Fourier fit features}   \\
\hline
a$_1$, .., a$_7$              & The amplitude of the Harmonics of the Fourier fit  \\
$\phi_1$, ..., $\phi_7$       & The phase shift of the Harmonics of the Fourier fit  \\
a$_{21}$, a$_{31}$, a$_{41}$     & The amplitude ratios of the Harmonics, (e.g., a$_{21}$ is a$_2$ over a$_1$, etc..) \\
$\phi_{21}$, $\phi_{31}$, $\phi_{41}$  & The phase shift ratios of the Harmonics (e.g., $\phi_{21}$ is $\phi_2$ over $\phi_1$) \\
\hline
\multicolumn{2}{c}{Group 4: Post classification features}   \\
\hline
Probability            & The Random Forest output probability \\
Template MSE         & Mean square error of the RRab template fit (Braga et al. 2019) \\
S/N                    & The signal to noise of the phased light curve as defined in Eq.~\ref{SN} \\
Nbins                  & Number of empty bins if the folded lightcurve is binned (in 20 bins) along the phase. \\

\hline\hline
\end{tabular}
\tablefoot{ 
 \\
$^1$ Anderson \& Darling 1952, Ann. Math. Statist. 23, 193. doi: 10.1214/aoms/1177729437 \\
$^2$ \citet{richards+11} \\
$^4$ \citet{kim11}  \\
$^3$ \citet{kinemuchi06} \\
}
\end{table*}

Several papers have used similar machine learning codes to identify RRL stars in large multi epoch catalogs,
and have defined different sets of features optimized for the characteristics of each specific survey \citep[e.g.][]{nun+15, elorrieta+16, medina+18, stringer+19, cabral+20, stringer+21, molnar+22, daza-perilla23}. After experimenting with different combinations of them, with a few minor modifications, we adopted the sets of features listed in Table~1.

The first group describes the properties of the time series photometry, independent of the period. These have the advantage of being robust against mistakes in the period determination. The second group describes the properties of the phase-folded light curve, and the third group includes parameters of the Fourier fit to the latter. One of the Random Forest outputs is the relevance of each feature, which in this case confirmed that the most relevant ones are -obviously- those in group 2.

In order to perform the Fourier fit to the light curves, we first run a Gaussian Process fit, using the python library {\it george} \citep{george_lib}. Then, we fitted a six-order Fourier series to the derived Gaussian process bins and calculated the relevant features as in \citet[their Fig.~1]{ogleIII}.

We activated the RF option to adjust the relative weight of the two classes of the training set (RRL and "other"), in order to artificially balance the number of stars in each of them. 

Finally, we fine-tuned the RF hyper-parameters, such as the maximum depth of the branches, the minimum number of objects in each leaf, the minimum number of objects required to split a node, and the number of random features to perform a split, by means of the 
{\it GridSearchCV} method of the sci-kit learn library. The latter allowed us to explore a grid of different combinations of these parameters, with the goal of maximizing the precision and the recall, described in the next section.


\subsubsection{Validation} 

In order to validate the output catalog, a standard 10-fold sampling was used, in which the training set was divided into 10 sub-samples of equal size, and then, iteratively, each of these 10 was used as a "test science sample" and classified using the other 9 as the training set. The result of this experiment demonstrated that the nominal precision of our classifier, when used with this training set, is 92\%, while the recall (completeness) is 97\%. This means that in the output catalog of stars with more than 50\% probability of being RRab, we expect that 92\% are indeed RRab, with the other 8\% being contamination from other variables. On the other hand, of the original sample of 18757 RRab in the training set, 97\% were correctly classified as such, while 3\% were incorrectly classified as other variables.

\begin{figure}
	\includegraphics[width=\hsize]{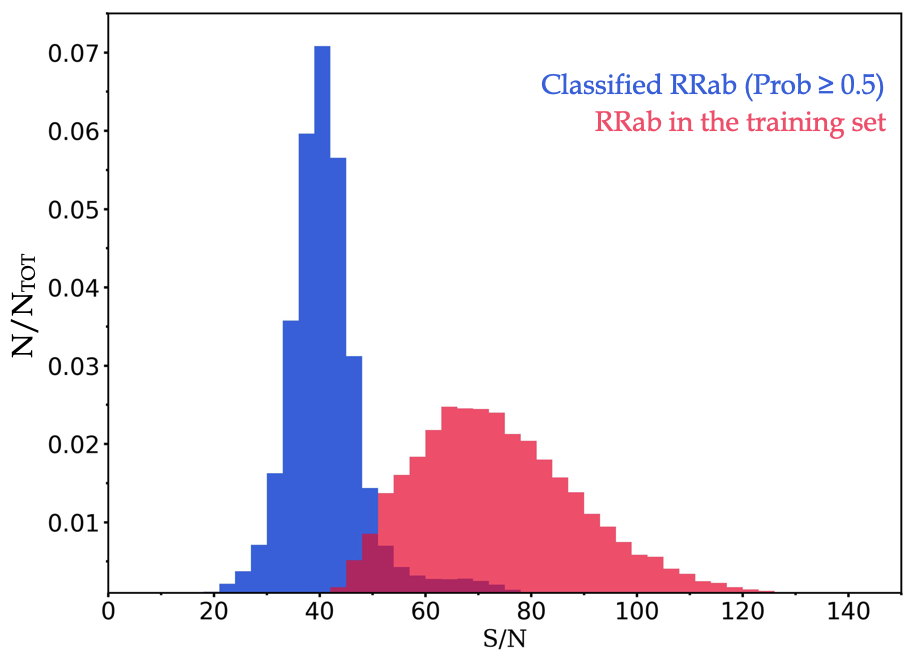}
 \caption{Normalized S/N distribution of the light curves for stars in the RRab training set (red) and in the science sample (blue).}
    \label{fig:SN}
\end{figure}

\subsubsection{Post Random Forest selection and visual inspection} 

When applied to the real science sample of about 10 million candidate variables, the classifier yielded a catalog of $\sim$490.000 stars with a probability higher than 50\% of being RRab variables. From visual inspection of several dozens of them, however, we concluded that the above quoted precision and recall were overestimated for the real sample. This is due to the fact that the training set includes variables detected relatively far from the midplane (see Fig. \ref{fig:sky_area}), in a relatively low extinction region of the sky, where the photometry is more precise. In the strip within $|b|<$2.5$^\circ$, instead, the extinction is larger. Variables are, therefore, fainter and not so well measured. In addition, the crowding is higher, making blends much more frequent. This can be verified by defining a "signal to noise" (hereafter S/N) for the light curves, defined as

\begin{equation}\label{SN}
S/N = A \frac{\sqrt{N}}{MSE}
\end{equation}

where A is the Amplitude defined in Table~1, N is the number of epochs, and MSE is the mean square error with respect to a Gaussian Process fit (sort of smoothing) of the light curve. Figure~\ref{fig:SN} shows that the S/N distribution of RRL in the training set is significantly higher than that of our science sample.
For the present analysis, we intentionally prioritized the purity (precision) of our catalog over the completeness (recall). Therefore, we blindly kept only the 9186 variables with $>$90\% probability of being RRab and S/N$>60$. In fact, as a safety check, we visually inspected and confirmed the light curves of a few hundred of those, randomly selected, but we also visually inspected all the candidates with probability $>$90\% but S/N$<$60.

The candidates with probability of being RRab lower than 50$\%$ were definitely rejected. Those with probabilities in the range 50$\%$--90$\%$ ($\sim$480,000 stars) were submitted to further cuts in order to select the most reliable ones, and the latter were all visually inspected. These post-classification selection criteria were based on the parameters defined in Table~1, Group~4, namely: S/N, MSE, Nbins and skewness. The latter cut is because a large skewness selects the light curves with the asymmetric saw-tooth shape, which is very characteristic of RRab, and it is hard to be confused with other kinds of variables. A star having a light curve with this shape allowed us to be more generous in its classification as RRab, even if other parameters were not optimal. The MSE, instead, is the mean square difference with respect to the fit of an RRL template from \citet{braga18}.

Specifically, we rejected all candidates with S/N$\lesssim $40 and Nbins$>$2 and skewness$<$0.04, i.e., the light curves that were noisy, or very unevenly sampled, or very symmetric. From all the others ($\sim$180,000), we extracted the group of $\sim$11,500 with Prob$>$80$\%$ and MSE$<$0.25, and the group of $\sim$4500 with 50$\%$$<$Prob$<$80$\%$ and MSE$<$0.20. These cuts, shown in Appendix~A were motivated by the attempt at isolating candidates as similar as possible as our golden sample, with Prob$>$90$\%$ and S/N>60 (see Appendix~A). 
These $\sim$16,000 candidate RRab variables were all visually inspected and eventually lowered to $\sim$4000.



Because we inspected the light curves of the majority of the RRab that made it to the final list, we trust that the present catalog optimizes purity. At the same time, since it relies upon visual classification as the final step, we realize the process leading to it is not completely objective nor repeatable. Yet, we believe that human classification is still the most reliable process, although the slowest. We designed the present RF classifier in order to lower the sample to be visually inspected, rather than as a tool to extract the final catalog in an objective way. We nonetheless provide the list of adopted features as an input for similar studies.

Note that the catalog released together with the present paper includes 3523 additional RRab from OGLE-IV, as explained in the next section, for a total of 16,488 variables.  We also release, in the online version, the values of the features for the 12965 variables classified here, together with their light curves.

\begin{figure}
\includegraphics[width=\hsize]{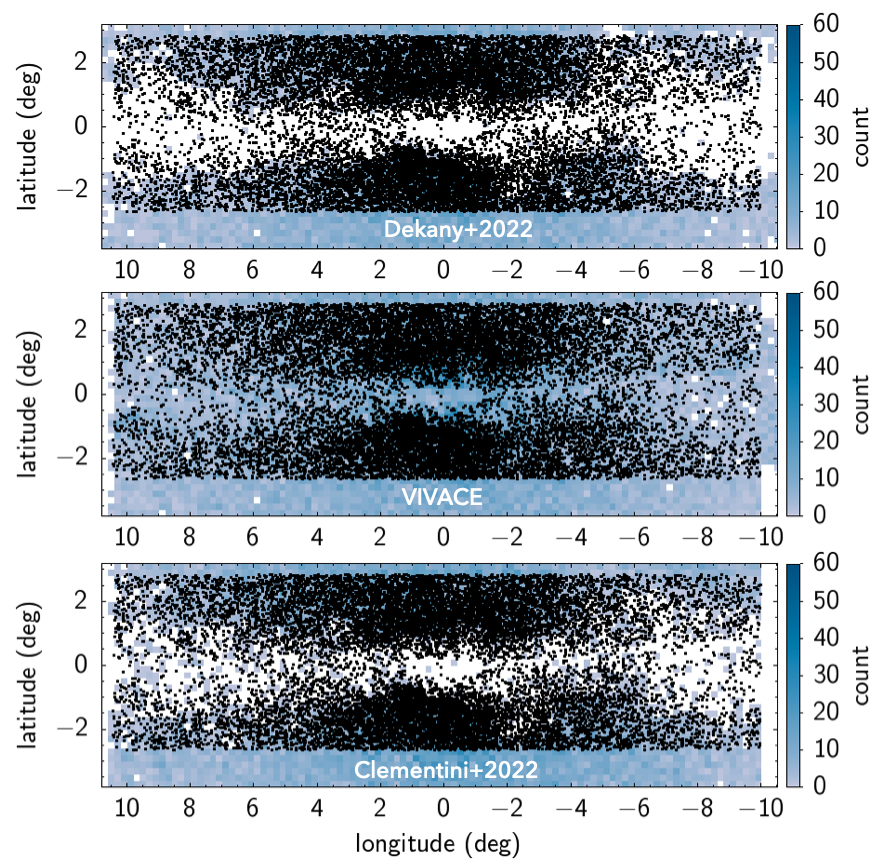}
 \caption{The footprints of the catalogs by \citet{dekany+22} (top), \citet{molnar+22} (middle) and \citet{clementini+22} (bottom) are shown in blue, with a density scale shown at the right of each panel. Black points show the galactic positions of the RRab in the present catalog, including the addition from OGLE-IV.}
    \label{fig:footprint}
\end{figure}

\section{Comparison with other RRab catalogs in the same region}
\label{sec:othercats}

In what follows we compare our catalog with other ones available in the literature and including the same region. The footprints of these catalog, compared to the present one, are shown in Fig.~\ref{fig:footprint}.
It is important to emphasize that none of these catalog is complete and free from contamination. Each of them has a different degree of these characteristics, which are both very difficult to assess a priori. For this reason, comparisons among different catalogs can help characterizing the performances of each, and perhaps choosing the most appropriate one, for future studies.
As discussed below, only for the OGLE-IV catalog we decided to add the RRab not found in our selection, but belonging to the same sky region. We did this only for OGLE-IV because this survey has a very fine cadence optimized for RRL detection and the S/N of the light curves is very high (Fig.~\ref{fig:SN}). These are the very high confidence variables that we adopted as training sample, therefore our analysis relies on the assumption that these are real RRab. The other catalogs listed below are based upon VVV data, like the present one. Therefore, although they reach closer to the Galactic midplane, they have sparser cadence and lower S/N light curves. Since our catalog has been optimized in favor of high purity, in contrast to larger completeness, we prefer to keep this characteristic and leave to the reader the option to add unmatched RRab from other public catalogs, if they consider it appropriate, based on their specific needs. 

\subsection{OGLE-IV}
There are 6881 RRab in common between our final catalog and OGLE-IV. On the other hand, 3523 OGLE RRab in the region of the sky analyzed here are not present in our catalog. With a flag (either "VVV" or "OGLE") we added those to the present release, as we trust that they are real RRab, and they might be useful targets for future follow-up studies. For 39 of them we cannot provide a mean J magnitude, therefore a reddening or a distance. We leave to the reader the decision whether to used them or not.
In the following analysis, we will show that the spatial distribution of these OGLE unmatched variables is biased toward close distances (see Sec.~\ref{sec:gcenter}). That is, they are seen preferentially at the near side of the Galactic bulge and disk, as expected, given that OGLE-IV is an optical survey conducted with a small telescope. Therefore, they will be excluded from the analysis of the spatialdistribution of bulge RRab stars.

\begin{figure}
	\includegraphics[width=\hsize]{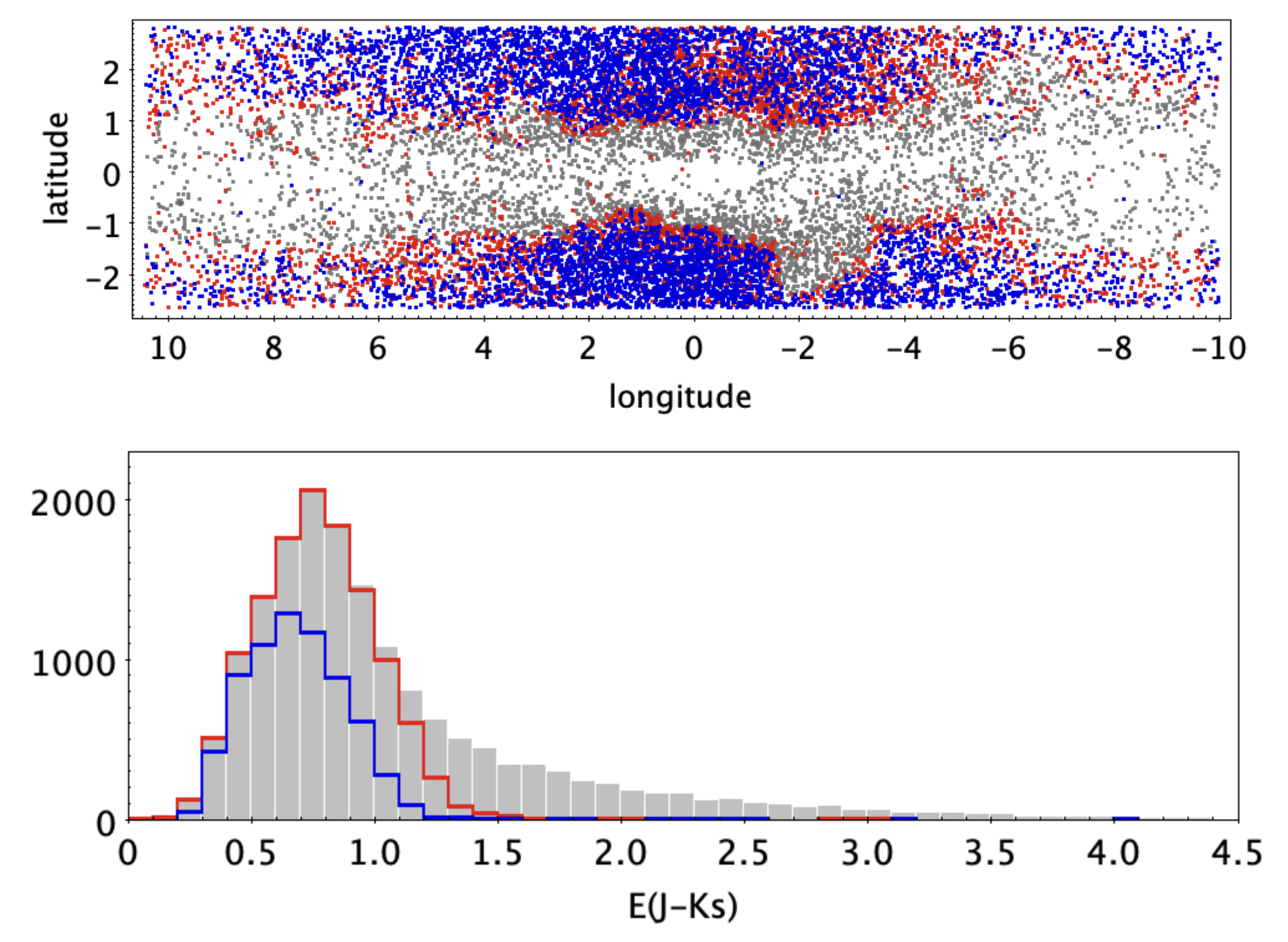}
    \caption{Top: Galactic position of each RRab in our catalog. In grey, we mark those present only in our catalog, including the OGLE-IV addition. In red, we mark those also present in the Gaia DR3 main catalog, and in blue those also identified as RRL, presented in \citet{clementini+22}. Bottom: Reddening distribution of the RRL in our catalog (grey), the fraction of them found in Gaia but not identified as RRL (red), and the fraction of them found and identified as RRL (blue). The color coding is the same in both panels}. 
    \label{fig:gaia}
\end{figure}

\subsection{D\'ek\'any et al. 2020}
Out of the 16,488 RRab in our catalog (12,965 from VVV + 3523 from OGLE-IV), only 11067 are in common with \citet{dekany+20}. 
We could not identify anything special in the properties of the unmatched 5421 variables: they have a similar distribution as all the others in amplitude, period, mean magnitude, probability, and S/N.

On the other hand, there are $\sim$1709 variables in the common region between \citet{dekany+20} and this study that were considered RRab in the former but are not present in our catalog. They are preferentially lower amplitude RRab ($<$Amp$>$=0.23 mag, versus $<$Amp$>$=0.28 mag for the whole D20 catalog\footnote{Note that the whole amplitude range for RRab in \ks is 0.1-0.5 mag, hence this difference is non-negligible.}), and lower S/N ($<$S/N$>$=65, versus $<$S/N$>$=88 for the whole D20 catalog), but have nothing special in either probability, period, or mean magnitude.

\subsection{VIVACE}
Only 903 RRab in our catalog ($\sim$6\%) are absent in the VIVACE catalog by 
\citet{molnar+22}. On one hand, there almost 8,000 RRab in the VIVACE catalog that are not in our sample, although they do lie in the common region of the sky. This is a rather large number that can be explained because the VIVACE catalogs have been created with the explicit intention of prioritizing completeness over purity. Therefore, the latter should be used as a more complete list of candidates whose real nature must be double-checked before drawing scientific results from it. On the other hand, our list offers a cleaner catalog, though definitely not complete, especially at latitudes very close to the midplane. 

\subsection{Gaia DR3}
There are 12,210 stars in common between the present catalog (including the addition from OGLE-IV) and Gaia DR3. Out of those, however, only 6816 were identified as RRL by Gaia \citep{clementini+22}, as shown in Fig.~\ref{fig:gaia}. The figure clearly shows that whether one of our variables is present in Gaia or not, depends almost exclusively on its extinction, affecting the quality of the Gaia optical photometry more than the VVV one. Indeed,  in the outer region, where Gaia can see through the interstellar dust, almost all our sources are measured by Gaia and identified as RRL. As shown by the histograms in the bottom panel of Fig.~\ref{fig:gaia}, the fraction of stars in our catalog detected by Gaia decreases with increasing \ejksnosp, as it does the fraction of the stars classified as RRL by \citet{clementini+22}. At low reddening (\ejks$<$1.1) the fraction of stars classified as RRL in our 
catalog but not in Clementini's is marginal. This validates our strategy of building a pure catalog, although certainly incomplete, especially in the region closer to the Galactic plane. 


\section{Metallicities, Reddenings, and Distances}
\label{sec:metdist}

RRL stars are excellent distance indicators because they follow a well-defined relation between the Luminosity and the Metallicity in the optical, and a Period Luminosity Metallicity (PLZ) in the near IR.  The first ingredients to calculate the distances are the mean apparent magnitudes, to be compared with the absolute magnitudes given by the PLZ. The mean magnitudes in the \ks band have been derived by means of the code by \citet{dekany+22}, which performs a Fourier fitting of the light curve, just like we did to derive the 3$^{\rm rd}$ group of features in Table~1. The advantage of using the code by \citet{dekany+22} is that it performs a bootstrap on the light curves, recalculates the mean by Fourier fitting at each iteration, and finally yields an error as the standard deviation of these determinations.
The mean magnitude in the J band is more complicated to derive because the light curves include only $\sim$10 points. The code by \citet{hajdu+18} fits a template light curve to these points and derives the mean magnitude and error on the template light curve. The typical errors on the mean magnitudes are $0.005$ and $0.060$ mag in the \ks and J band, respectively.

\begin{figure}
	\includegraphics[width=\hsize]{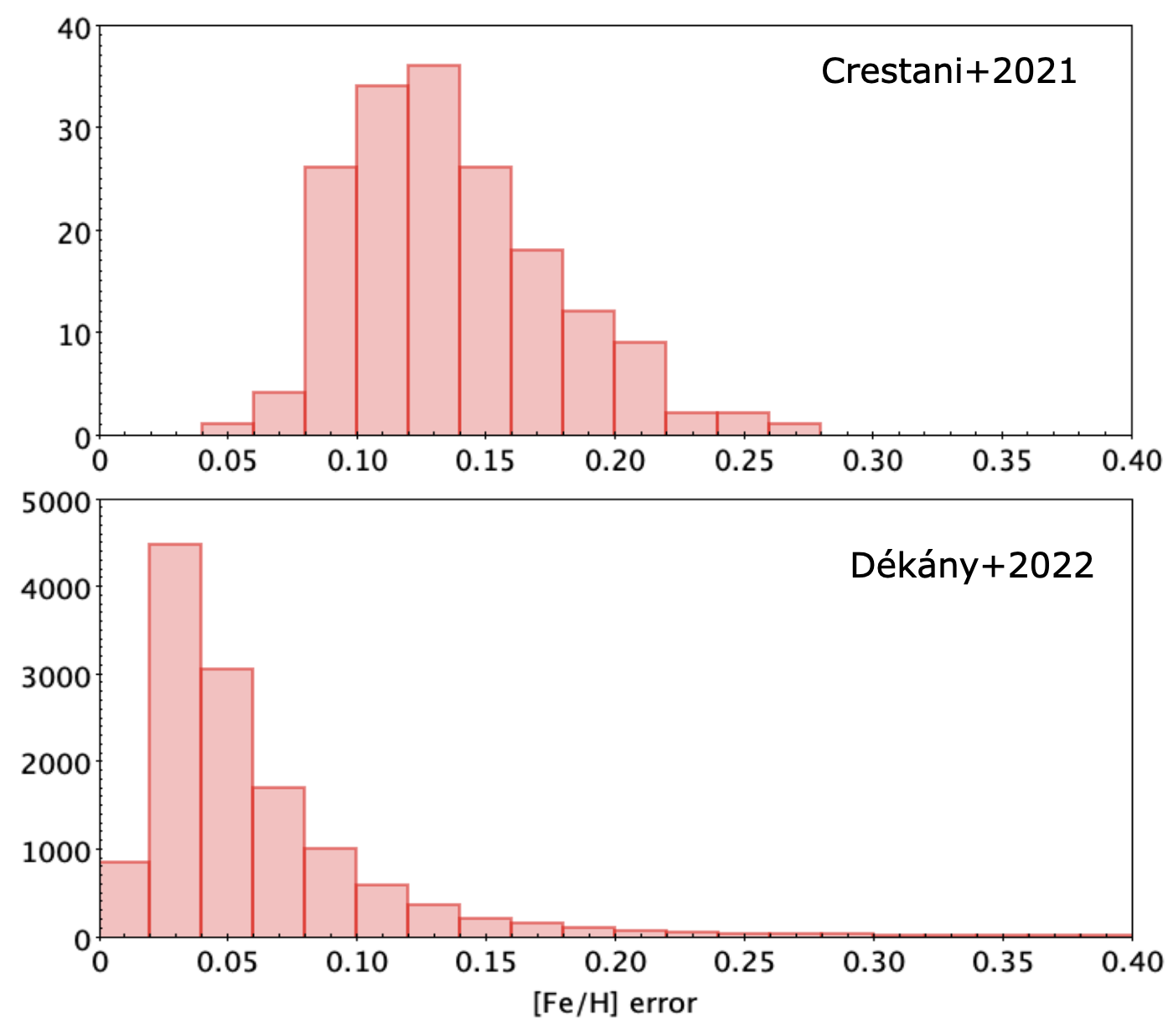}
    \caption{Top panel: distribution of the estimated errors in the [Fe/H] as quoted in \citet[their Table~2]{crestani+21}. The latter work is based on high dispersion spectra (R=35,000) for 143 RRL. Bottom: distribution of the estimated errors in the [Fe/H] values inferred by \citet{dekany+22} from the shape of the light curve by means of a machine learning algorithm trained (indirectly) on the spectroscopic measurements by \citet{crestani+21}.} 
    \label{fig:errFeH}
\end{figure}

The metallicities for our RRab sample were estimated from the shape of their light curve, following the prescriptions by \citet{dekany+22}. The corresponding machine learning code available online also yields metallicity errors, which are peaked at $\Delta$[Fe/H]=0.025 dex with a median value of $\Delta$[Fe/H]=0.045 dex (Fig~\ref{fig:errFeH}, bottom). Nonetheless, the uncertainties on the derivation of these photometric metallicities should be at least slightly larger than the errors on the metallicities of the training sample, i.e., the \citet{crestani+21} sample, whose error distribution is shown in the top panel of Fig.~\ref{fig:errFeH}. In order to correct this inconsistency, in what follows, we will adopt a constant $\Delta$[Fe/H]=0.3 dex, as a conservative error estimate, for all the variables in our catalog (see Sec.~\ref{sec:dist-err}).

Reddenings to individual variables have been estimated from the observed \jks colors. From the distribution of RRab members of the globular cluster $\omega$ Centauri \citep{navarrete15}, we found that they are distributed around a mean intrinsic color of (\jksnosp)$_0$=0.26 with a standard deviation of 0.03 mag, and an error on the mean of 0.003 mag. The above quoted dispersion of 0.03 magnitudes is negligible compared with the large range of reddenings observed towards the Galactic midplane \citep{surot20}, however, the effect of this approximation will be discussed in Sec.~\ref{sec:dist-err}.
We then calculated the color excess of individual variables as the difference between the observed \jks color and the intrinsic value of 0.26 mag. 
In order to estimate the effect of reddening on the mean \ks magnitudes, and therefore on the distances, we adopted the total to selective extinction ratio derived by \citet{minniti+20}, from a sample of classical Cepheids at the far side of the Galactic disk, i.e., appropriate for the $b=0^\circ$ direction, which is:
\begin{equation}
{\rm A_{K_S}=(0.465\pm0.022)\times E(J-K_S). }
\end{equation}

\subsection{Choosing the appropriate PLZ relation}
\label{sec:plz}

\begin{table}
\caption{The PLZ relations examined in the present work.}
\tiny
\label{table:PLZ}      
\begin{tabular}{l}
\hline \hline
\citet{muraveva18}: \\
~~ M$_{\rm Ks}$ = ($-$2.58$\pm$0.20) $\times$ $\log$ P + (0.17$\pm$0.03) [Fe/H] $-$ (0.84$\pm$0.09)   \\
\\
\citet{neely+19}: \\
~~ M$_{\rm Ks}$ = ($-$2.45$\pm$0.28) $\times$ ($\log$ P + 0.3) + (0.20$\pm$0.03) $\times$ ([Fe/H]+1.36) \\
~~~~~~~~~~~ $-$ (0.37$\pm$0.02) \\
~~ M$_{\rm J}$  = ($-$1.91$\pm$0.29) $\times$ ($\log$ P + 0.3) + (0.17$\pm$0.03) $\times$ ([Fe/H]+1.36) \\
~~~~~~~~~~~ $-$ (0.14$\pm$0.02) \\
\\
\citet{bartolomeo+23}:   \\
~~ M$_{\rm Ks}$ = ($-$3.10$\pm$0.39) $\times$ ($\log$ P+0.25) + (0.080$\pm$0.029) $\times$ ([Fe/H]+1.5) \\
~~~~~~~~~~~ $-$(0.404$\pm$0.016) \\
~~ M$_{\rm J}$  = ($-$3.09$\pm$0.52) $\times$ ($\log$ P+0.25) + (0.070$\pm$0.030) $\times$ ([Fe/H]+1.5) \\
~~~~~~~~~~~ $-$(0.147$\pm$0.020) \\
\\
\citet{bhardwaj+23}: \\
~~ M$_{\rm Ks}$ = ($-$2.39$\pm$0.03) $\times$ $\log$ P + (0.18$\pm$0.02) $\times$ [Fe/H] $-$ (0.81$\pm$0.03) \\
~~ M$_{\rm J}$  = ($-$1.97$\pm$0.04) $\times$ $\log$ P + (0.20$\pm$0.01) $\times$ [Fe/H] $-$ (0.46$\pm$0.02) \\
\\
\citet{prudil+23}: \\
~~ M$_{\rm Ks}$ = $-$2.342 $\times$ $\log$ P + 0.138 $\times$ [Fe/H] $-$ 0.801 \\
~~ M$_{\rm J}$  = $-$1.799 $\times$ $\log$ P + 0.160 $\times$ [Fe/H] $-$ 0.378 \\
\hline
\end{tabular}
\end{table}

Several PLZ relations are available in the literature, allowing us to derive the absolute magnitudes of individual RRab variables from their observed periods (PL) and periods plus metallicities (PLZ). With the absolute magnitude and the extinction estimated above, the distance can be derived from the definition of distance modulus:

\begin{equation}
d = 10^{1+0.2(K_S-A_{K_S}-M_{K_S})}.
\end{equation}

For the present work we explored several recent PLZ relations, whose functional form are quoted in Table~\ref{table:PLZ} for the reader's convenience.
Initially, we selected the PLZ relation from \citet[][hereafter M18; their Table 4]{muraveva18}, derived based on Gaia DR2 parallaxes for a sample of MW field RRL with metallicities in the same range as those in our sample. 

\begin{figure}
	\includegraphics[width=\hsize]{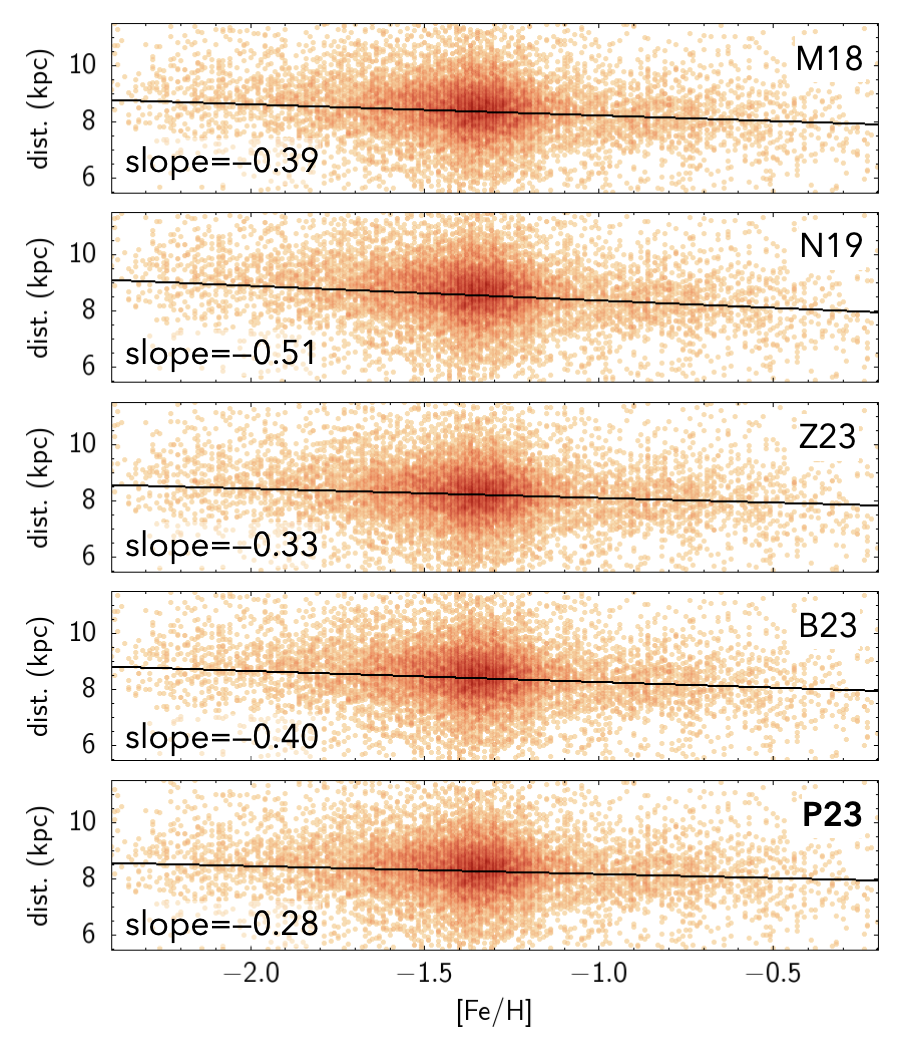}
    \caption{The trend of the heliocentric distances to individual RRab in our catalog versus their photometric metallicity. From top to bottom, the adopted PLZ are those by \citet{muraveva18, neely+19, bartolomeo+23, bhardwaj+23, prudil+23}. A linear fit highlights the presence of a slope, whose value is written in the labels, that is minimized for the PLZ by \citet{prudil+23}, which we select as the optimal one for the present work.} 
    \label{fig:d_feh}
\end{figure}

When analyzing the 3D distribution of the Galactic component traced by RRab variables, however, we noticed that metal-rich RRab were placed systematically closer to the Sun, with respect to metal-poor RRab. Indeed, there is a trend between the distance obtained from the M18 relation and the photometric metallicity used as input (Fig.~\ref{fig:d_feh}; top panel). Such a trend is certainly artificial, although at this time we cannot be sure whether it is due to the photometric metallicities being wrong or to the metallicity coefficient of the M18 PLZ being too large. In the first case, we would be attributing a different metallicity to two groups of variables that are, on average, identical. In this case, by means of the PLZ, we would be applying a correction factor that is not appropriate, artificially producing the observed segregation.
We notice that a possible reason why the photometric metallicities derived here could have a large error is that our light curves are relatively noisy (Fig.~\ref{fig:lc}), potentially resulting in the high-order Fourier coefficients being less reliable than they are for the RRab of the outer bulge, used by \citet{dekany+22} to derive the calibration.
The same effect would be seen if the estimated metallicity is correct but the metallicity coefficient of the PLZ is overestimated. With our data alone, we cannot distinguish between these two alternatives. Therefore, we decided to explore alternative relations available in the literature. 

\begin{figure}
	\includegraphics[width=\hsize]{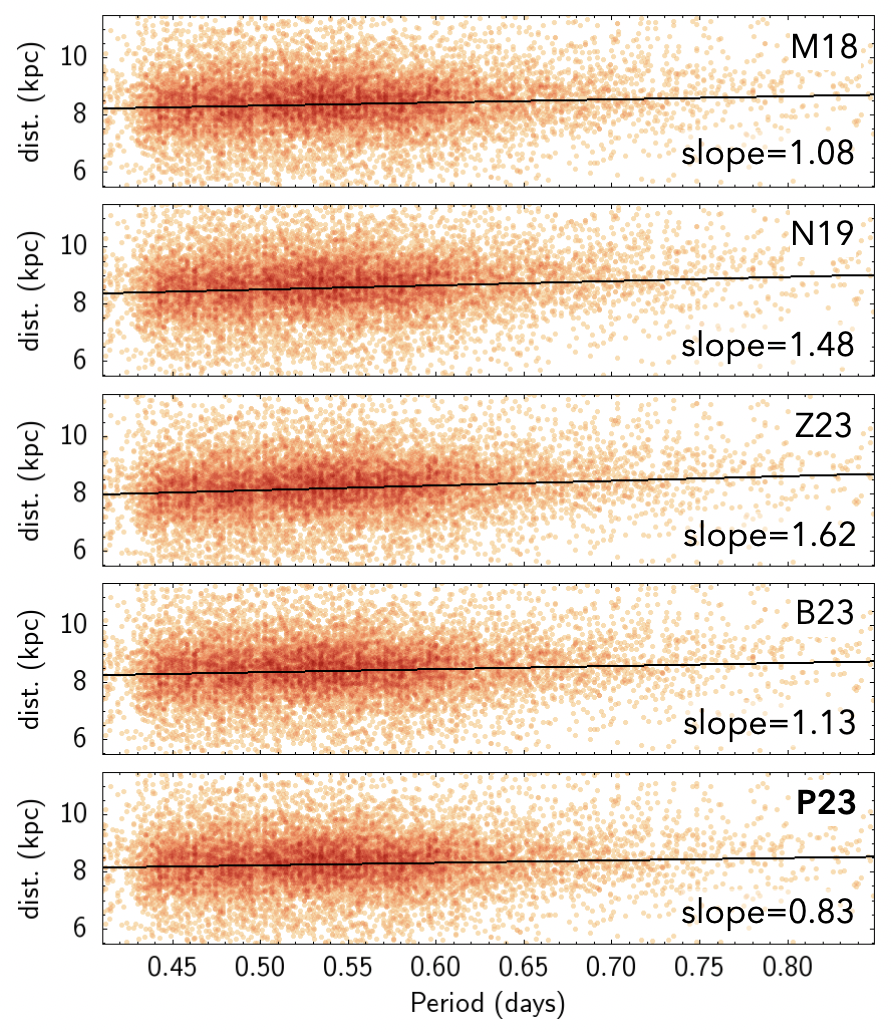}
    \caption{Same as Fig.~\ref{fig:d_feh}, but now distances are plotted against period. A linear fit highlights the presence of 
     a slope, that should not be there, whose value is written on the labels. The slope is smaller for the PLZ by \citet{prudil+23}, finally used in the present analysis}. 
    \label{fig:d_p}
\end{figure}

The PLZ provided by \citet{neely+19} showed a similar problem, actually larger (Fig.~\ref{fig:d_feh}).
At the time of writing this manuscript, new PLZ relations in the near IR were published by \cite[][hereafter Z23]{bartolomeo+23}, based on Gaia DR3 parallaxes for a sample of 28 local RRL variables, by \citet[][hereafter B23]{bhardwaj+23}, based on 964 RRL in 11 globular clusters and 346 field RRL with Gaia parallaxes, and by \citet[][hereafter P23]{prudil+23}, based on $>$100 field RRL stars with Gaia parallaxes. We applied all of them to the RRab in our catalog and checked against a residual trend of the derived distance with metallicity. The result is shown in the different panels of Fig.~\ref{fig:d_feh}, together with a linear fit, revealing that the most appropriate relation for our metallicity scale is the PLZ by P23.

For all the above relations, we also checked against residual (spurious) trends against period, as shown in Fig.~\ref{fig:d_p}. These plots confirm that the PLZ by P23 is also the one that minimizes the trend with periods, although without completely removing it.

In order to calculate the reddening suffered by our RRab, we adopted a constant intrinsic color
(\jksnosp)$_0$=0.26 mag, independent from the period. Because the PLZ in the literature have a different coefficient for the period, in J and K$_{\rm s}$, they predict a small dependence of the intrinsic color upon the period, that we have neglected in this work. The coefficients between the intrinsic (\jksnosp)$_0$ and $\log$~P range from $-$0.01, for the PLZ by \citet{bartolomeo+23}, to $-$0.54 for the PLZ by \citet{prudil+23}, respectively.
In order to verify that neglecting this dependence is not the reason for the trends of distances with period
shown in Fig.~\ref{fig:d_p}, we calculated the differences between the adopted color (\jksnosp)$_0$=0.26 mag and the color predicted by each of the PLZ explored above. The difference is small, reaching 0.05 mag at the edges of the period distribution, and it translates to a difference in A$_{\rm K_s}$ of $\sim$0.025 mag.
The difference goes in the sense that, at low periods, the distance calculated here is larger than the one calculated by assuming the color predicted from the PLZ, and at high period the distance calculated here is smaller. In other words, had we assumed the expected color derived from the PLZ, instead than a fixed value, all the slopes in Fig.~\ref{fig:d_p} would have been even larger.


The above exercise demonstrates that a new PL relation could be derived here, by using the present catalog alone.
Indeed, we could derive a zero point by imposing that the mode of the projected distance coincides with the distance to the GC derived by the GRAVITY experiment and a slope that flattens the trend between distance and periods. We anticipate that the period slope that we would find by this method is $-2.7$, however, we postpone a more extensive discussion to a forthcoming paper, where we will properly include errors. The reason is that, as we will discuss below, errors on the distances are key to estimate the shape parameters traced by RRL.  The reason why the PLZ by M18, B23, and Z23 perform well in the context where they were derived but not in the inner Galactic bulge remains to be fully understood.

\subsection{Errors on the distances}
\label{sec:dist-err}

It is important to characterize the error in the distance to individual stars because it will impact the inferred shape. To put it simple, because the error on the distances will always be larger than the
error on the galactic coordinates, the distribution we will infer will always be artificially elongated
(or extra-elongated) in the direction of the line of sight. A proper correction for this effect requires
a knowledge of the distance errors as accurate as possible.

Besides the errors in the coefficients of the PLZ and in the extinction law quoted above, we estimated the errors in the periods to be negligible since our periods and the one derived by OGLE-IV for the stars in common, agree to the third decimal digit. 

The statistical errors on the distances were calculated by applying the standard error propagation on equation~4. The errors on the mean magnitude of each RRab were derived from the code by \citet{dekany+20} for the \ks band and \citet{hajdu+20} for the J band, as mentioned in Sec.~\ref{sec:metdist}.
The error on the extinction comes from the propagation of the latter into equation~3. As explained at the beginning of this section, the error on the metallicity has been assumed to be $\Delta$[Fe/H]=0.3 dex for all the stars. 

\begin{figure}
	\includegraphics[width=\hsize]{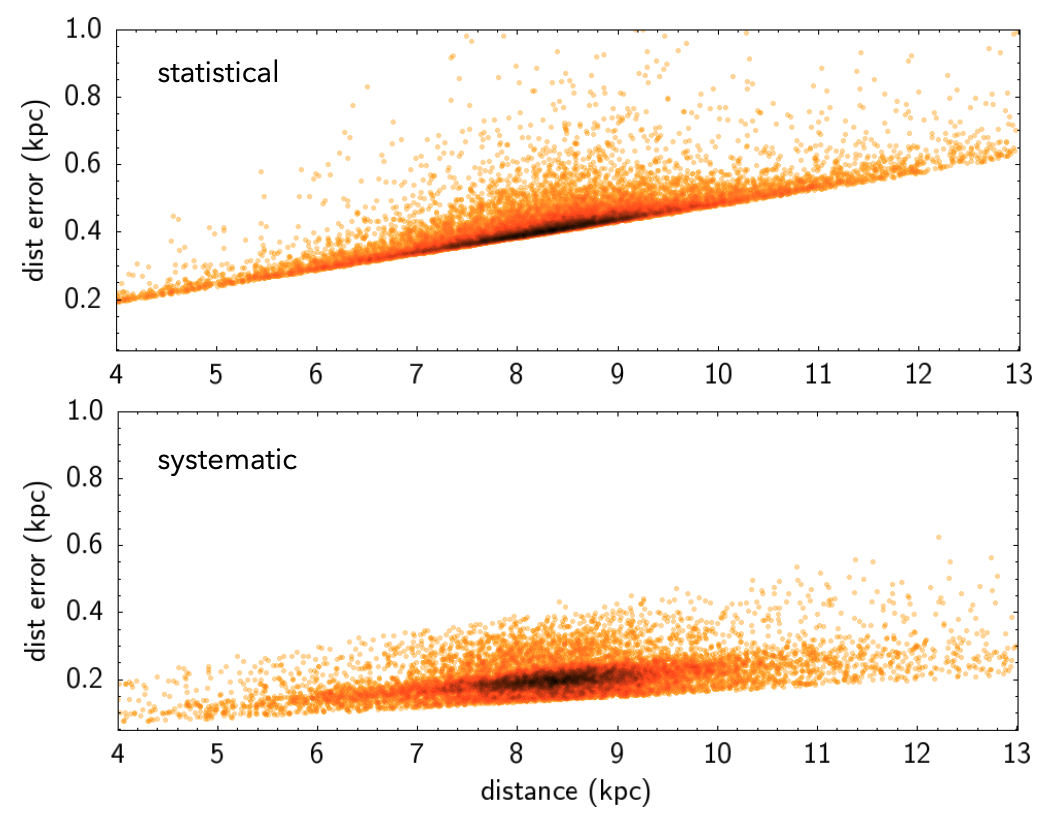}
    \caption{Top: Distribution of the RRL distances and their statistical errors, derived adopting the PLZ by P23. Bottom: Our estimate of the systematic error associated to the distance measurements, obtained as the standard deviation of the distances derived adopting the five different PLZ examined here.}
    \label{fig:err_d}
\end{figure}

We noticed that the difference in distance obtained when adopting one or the other PLZ, shown in the bottom panel of Fig.~\ref{fig:err_d}, is not negligible compared to the statistical error calculated as explained above, and shown in the top panel of the same figure. Therefore, we considered the former as an estimate of the systematic error that should be added to the global error budget for a realistic quantification of the total distance uncertainty. Accordingly, in what follows we will consider as the total error on the distance of each star, the square sum of the statistical error plus a linear fit to the systematic error as a function of distance.

It should be noted that the spread among the distances obtained with different PLZ is not, strictly speaking, a systematic error, i.e., it is not the same for all the stars. However, it is a source of error that it is not accounted for in the quoted uncertainty in the PLZ coefficients. Indeed, the difference between the distance derived from the M18 or the Z23 relations (which are at the two extremes) is larger than the formal, statistical error derived by applying the error propagation on each 
of the two. The way we quantify this systematic here is just an estimate, as we examined only five of the available PLZ in the literature. 

For practical purposes, it does not affect our results, as the total error derived as explained above is just slightly larger than the statistical error. Nonetheless, we consider it important to highlight the fact that the errors on each of the ingredients of the PLZ seem to underestimate the total uncertainty on the derived distances.


\section{Distance to the Galactic center}
\label{sec:gcenter}

With the distances calculated in Sec.~\ref{sec:plz}, the first sanity check to be done is to derive the distance to the GC (R$_{\rm GC}$), as already done by several authors using distance indicators. Recent determinations using RRL are found in \citet{griv+20},  based on OGLE-IV RRL, and in \citet{majaess+18} using RRL from VVV at |b|$>4^\circ$. A review of results by different authors, derived by means of variable stars, is given by \citet{bobylev+21}, who recommend a combined value of R$_{\rm GC}$=8.1$\pm$0.1\,kpc. On the other hand, a review of results using different methods is given by \citet[][their Fig.~1]{leung+23}. 

\begin{figure}
	\includegraphics[width=\hsize]{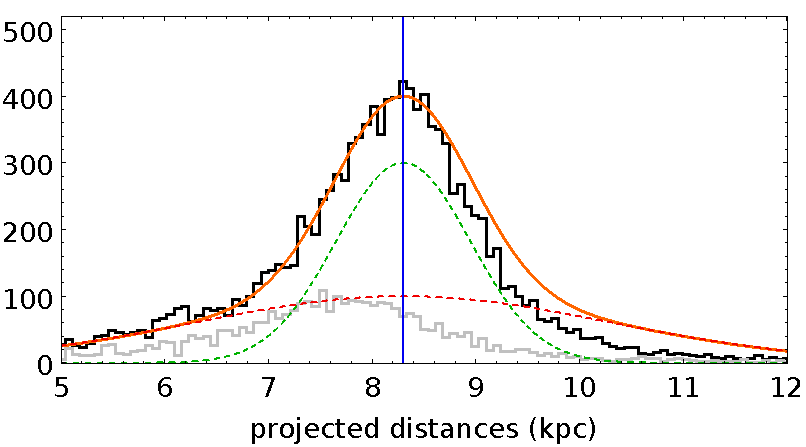}
    \caption{Histogram of the distances to individual variables, projected along the Sun-GC direction. The black histogram shows the 12,965 variables identified by our algorithm, while the light grey histogram shows the 3521 RRab that were added from the OGLE-IV catalog. The latter have a clear bias towards more nearby variables, and therefore, they are excluded from the analysis. The green dotted Gaussian represents the projected bulge density distribution, while the red dotted one refers to the disk. Their sum, reproducing the observed distribution, is the orange thick line. The latter is 
    centered at R$_{\rm GC}$=8.30\,kpc (blue vertical line).}
    \label{fig:GCdist}
\end{figure}

Here, the distances to individual variables were projected to the Galactic plane and then to the Sun-GC direction by means of their galactic latitude and longitude, respectively. A simple histogram was then constructed, including a correction factor for the so-called "cone effect", i.e., the fact that a fixed field of view maps larger volumes at larger distances, therefore the same distance bin, at larger distances, contains more stars than at shorter distances. The result is shown in Fig.~\ref{fig:GCdist}. The distribution peaks at R$_{\rm GC}$=8.30\,kpc, and it is qualitatively compatible with the sum of two Gaussians, one representing the disk and the other the bulge. As expected, our catalog is more complete on the near side of the disk (and bulge) than on the far side. The grey histogram shows the RRab from the OGLE-IV catalog that were not identified by our algorithm and were added later. Their distribution is much more skewed toward short distances due to the use of optical bands and the smaller telescope. Therefore, although we include these stars in the catalog, they are excluded from this and the following analysis.

The derived distance to the GC is consistent both with previous results with similar methods and with the much more precise result of R$_{\rm GC}$=8.275$\pm$0.034\,kpc obtained by modeling the orbit of the stars around the Galaxy supermassive black hole \citep{gravity+21, gravity+22}. More than claiming a new determination, we consider this qualitative analysis as a sanity check against important biases such as the one seen in the OGLE-IV variables. Indeed, it is reassuring that the selected PLZ, in addition of minimizing the spurious dependence of the distances upon metallicity and period, also has a zero point that sets the center of the RRab distribution at the correct position for the GC.

\section{The shape of the bulge RR Lyrae parent population}
\label{sec:3D}

\begin{figure}
    \includegraphics[width=\hsize]{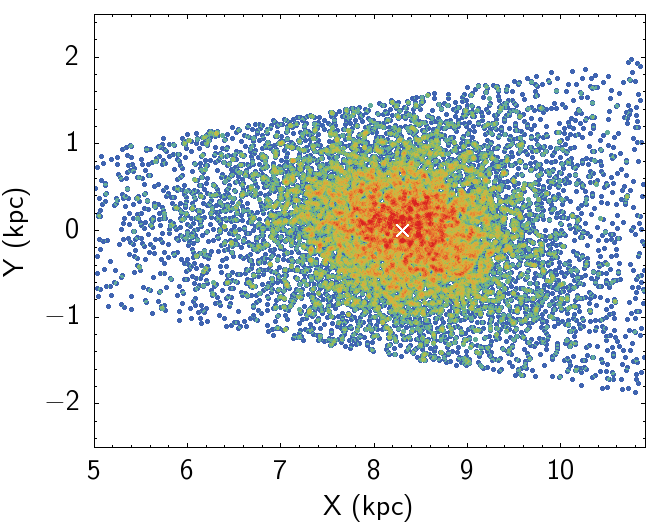}
    \caption{Cartesian position of individual RRab in our catalog, projected onto the Galactic plane X,Y. The white cross marks the GC, and the Sun is at (0,0).}
    \label{fig:XY_raw}
\end{figure}

In order to infer the 3D distribution of RRab around the GC, the Galactic coordinates of the stars, together with their distances, were converted into standard Cartesian (X,Y,Z) positions by means of the standard formulae: \\

\hspace{2cm} ${\rm X = d ~ cos}\,l ~ {\rm cos}\,b$  \hfill

\hspace{2cm} ${\rm Y = d ~ sin}\,l ~ {\rm cos}\,b$  \hfill 

\hspace{2cm} ${\rm Z = d ~ sin}\,b$ . \hfill \break

\begin{figure}
    \includegraphics[width=\hsize]{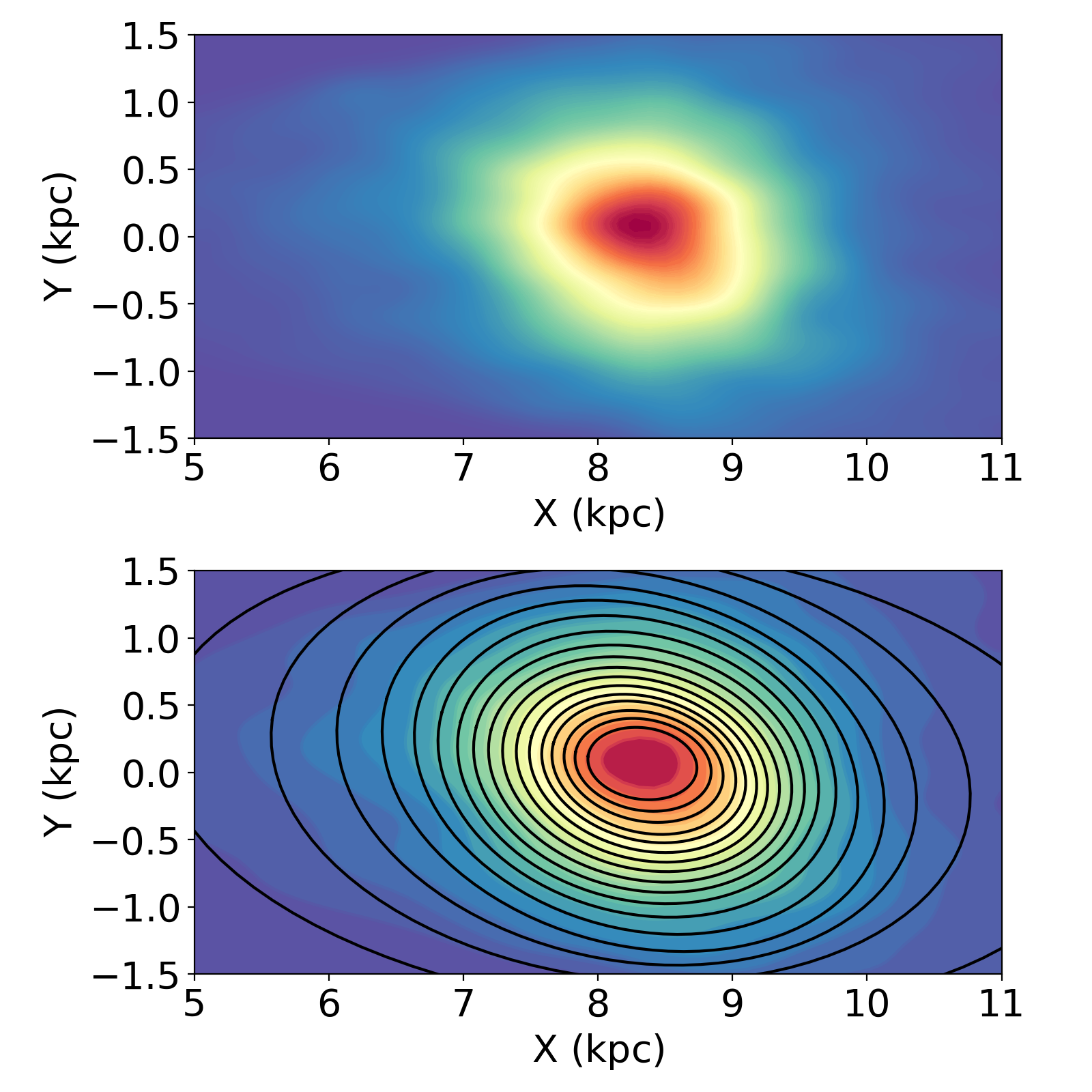}
    \caption{Top: Two dimensional KDE of the distribution of the RRab, projected onto the Galactic plane X,Y. Top: Color shades show density contours. Bottom: Ellipses (black lines) are fitted to several selected density contours. }
    \label{fig:KDE}
\end{figure}

The resulting (X,Y) distribution is shown in Fig.~\ref{fig:XY_raw}, with the Sun located at the (0,0) position. A Kernel Density Estimator (KDE) is then applied to this distribution to highlight its shape and derive its parameters (ellipticity and inclination). The top panel of Fig.~\ref{fig:KDE} shows the KDE with minimum smoothing. Color shades show density contours. Because different contours have different shapes, which are also highly variable with the adopted smoothing parameters, we decided to fit ellipses to several contours (bottom panel) and combine their ellipticity and inclination angle in order to provide values representative of the whole distribution. Accordingly, the median inclination angle between the major axis of the ellipses in Fig.~\ref{fig:KDE} and the Sun-GC direction is $\phi$=15.1 degrees, with a standard deviation of 3.04, while the median axis ratio b/a is 0.69$\pm$0.05. Therefore, the derived median ellipticity is 
\[
{\rm e}=\sqrt{1 - \frac{b^2}{a^2}} = 0.72 \pm 0.07
\]

\begin{figure}
	\includegraphics[width=\hsize]{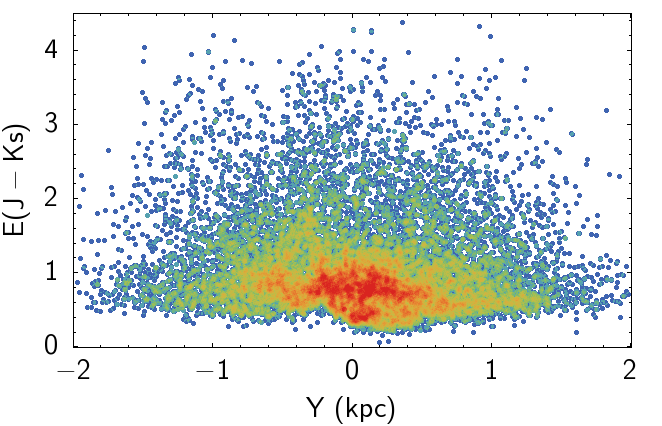}
    \caption{Reddening versus Y coordinate for all the RRab in our catalogue. It is clear that the reddening distribution is not symmetric with respect to the Y=0 direction, therefore its treatment can have an effect on the derived shape parameters traced with RRL. }
    \label{fig:EJK_Y}
\end{figure}




\begin{figure}
	\includegraphics[width=\hsize]{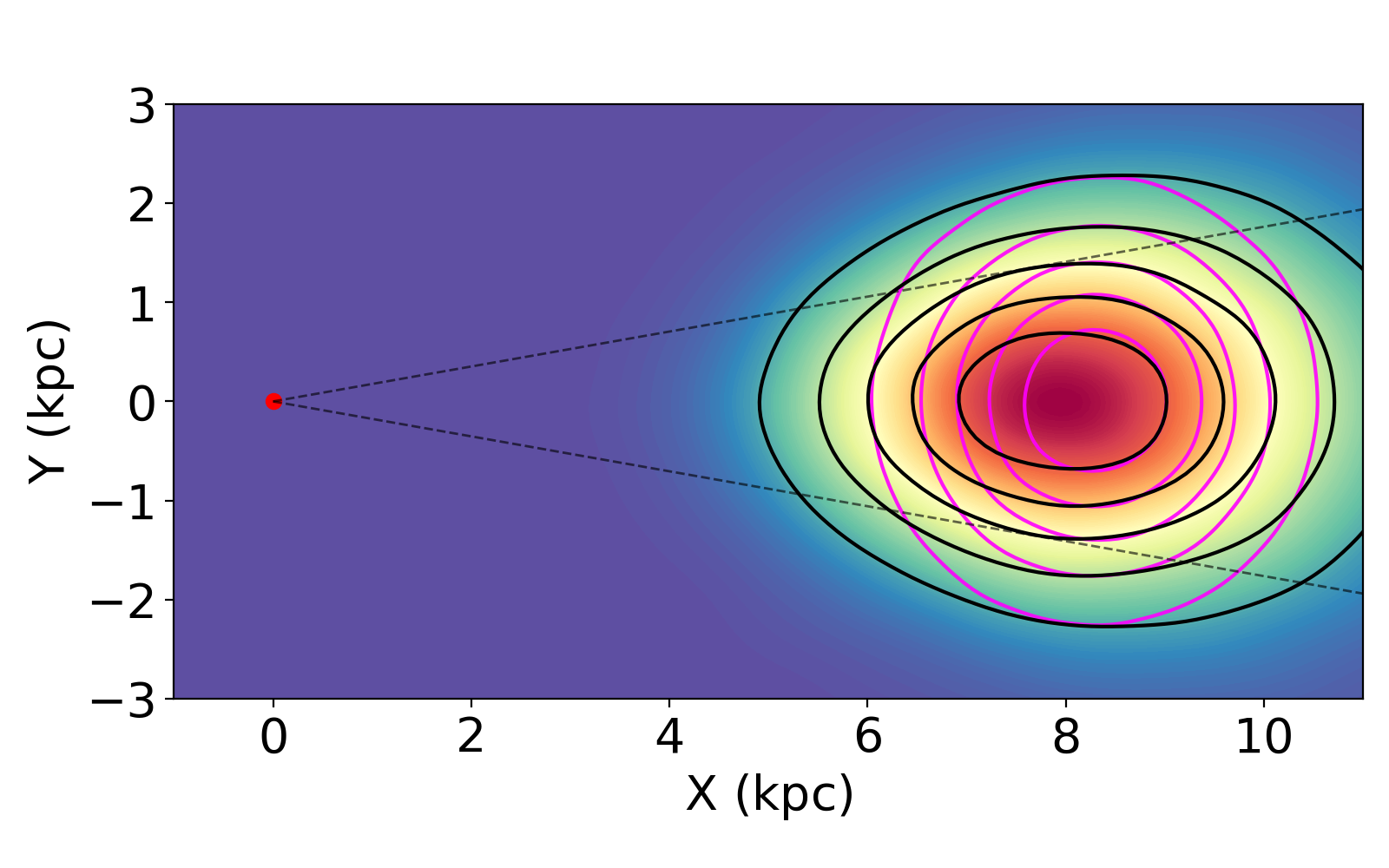}
    \caption{Qualitative representation of the effect of observational errors on the shape of the observed distribution of any sample of tracers. A perfectly spherical distribution, shown by the pink contours, is converted into an egg-shaped distribution (black contours) by the effect of the errors on the distances,
    which are always larger than the errors on the galactic coordinates. This would imply that a bar would be measured with a lower inclination angle with respect to the Sun-GC direction and a larger ellipticity.}
    \label{fig:huevo_did}
\end{figure}

The shape parameters derived above suffer from two biases, which we discuss below. 

First, the derived parameters would be correct under the hypothesis that the RRab completeness function is flat across the cartesian X,Y plane. That is, the catalog includes the same fraction of observed/real stars in every pixel of the plane in Fig.~\ref{fig:XY_raw}, so that the observed density is representative (though lower) of the true one. Unfortunately, there is no way to verify that. Even forcing a symmetry in the number of RRab at positive and negative longitudes would be inappropriate because the presence of a bar with different elongation parameters would result in a different degree of observed asymmetry in the density of tracers. In addition, and much more important, we know that the extinction is not uniform, across the sky. 
Although stochastic in nature, the specific reddening distribution of stars in the MW shows some degree of asymmetry between positive/negative projected coordinate Y, as shown in Fig.~\ref{fig:EJK_Y}. As a consecuence, the completeness of RRL is not symmetric with respect to the Y=0 direction, and, more importantly, the treatment of extinction may have an impact on the derived shape parameters, as it does not affect positive and negative Y in the same way. It is important to be aware of these effect, although we could not devise a way to correct for them.

As a second observational bias, we must take into account the effect of errors on the observed distribution. Indeed, because the errors in distance are significantly larger than the (negligible) errors in the Galactic coordinates, the observed distribution will always be more elongated along the line of sight compared to the real one, and the apparent inclination angle will be smaller \citep[c.f.,][]{hey+23, vislosky+24}. In order to qualitatively illustrate this effect, we show in Fig.~\ref{fig:huevo_did} how a typical error distribution would distort a perfectly spherical initial distribution (grey contours). Because the distance errors stretch the distribution along the line of sight, and because the latter is a cone, the observed distribution is egg-shaped, as shown by the black contours and color shading. This shape may be easily mistaken with an elongated bar, and parameters would be derived, that would not represent the original distribution. A proper quantification and correction of this effect requires proper modeling of the errors in distance and, therefore, depends on the sample. 

\begin{figure}
    \includegraphics[width=\hsize]{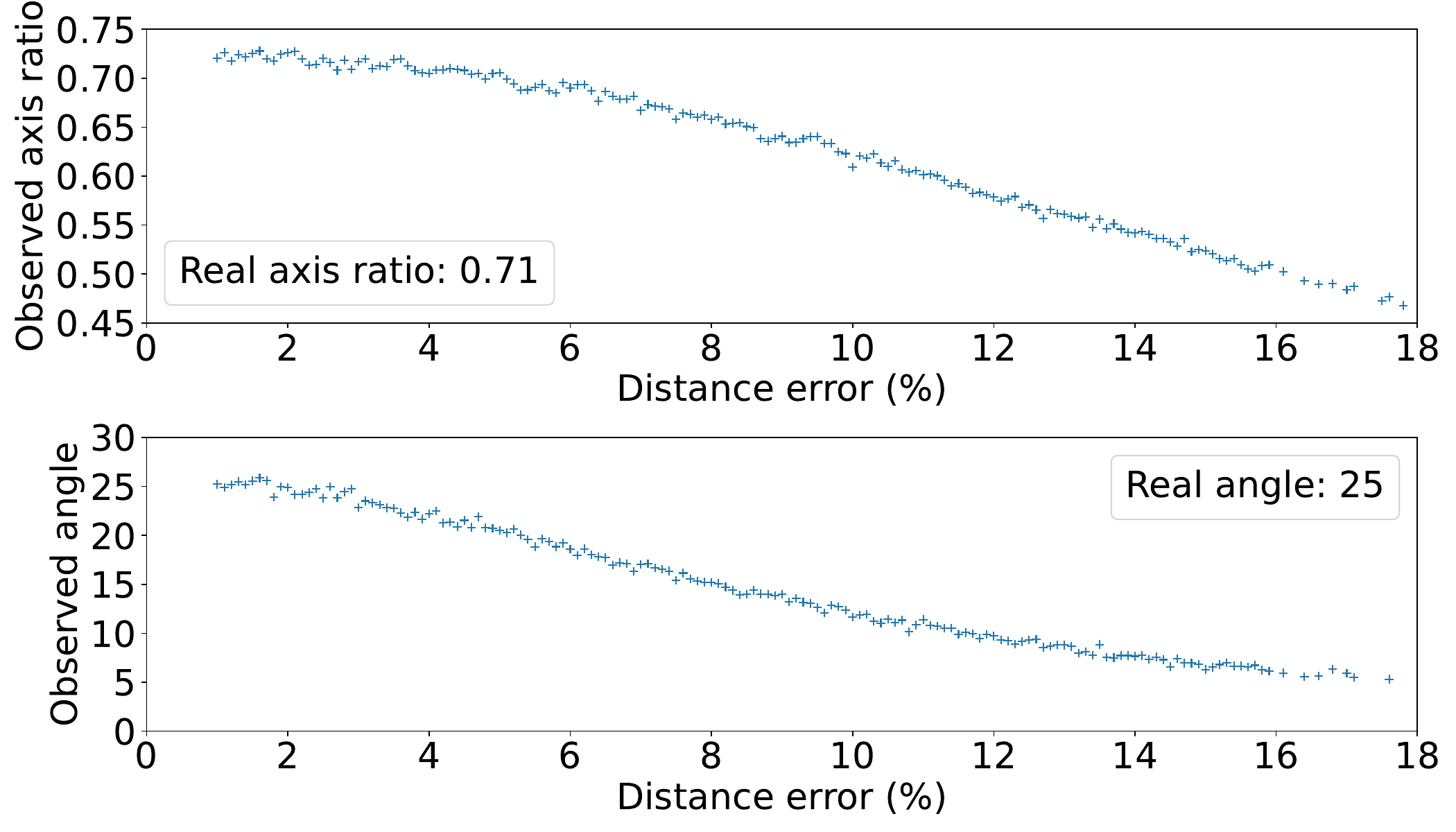}
    \caption{Observed axis ratio (top) and inclination angle (bottom) as a function of the percent error on the distance for a fixed input value of both parameters, as indicated by the labels. The simulation shows that at larger distance errors, the observed bar is more elongated than its true counterpart, and its observed major axis tends to align with the Sun-GC direction.}
    \label{fig:sim_huevo}
\end{figure}

We performed a simulation whereby an input mock distribution with a given inclination angle and axis ratio was affected by the error distribution shown in Fig.~\ref{fig:err_d}, and the observed parameters recovered. We iterated on the input distribution until we recovered the parameters actually observed in the present sample. In our case, we found that the difference between the input parameters and the observed ones is very small, within 1 degree for the inclination and negligible for the axis ratios. Nonetheless, shapes derived from other samples might be affected by larger errors. In order to provide the reader with an estimate of how important this effect can be for other tracers, 
we provide here a simulation of how a bar with axis ratio b/a=0.71 and inclination angle of $\phi$=25 degrees would be observed at increasing fractional distance errors (Fig.~\ref{fig:sim_huevo}). The input parameters adopted here have been chosen to be close to the typical values found in the literature, for the Galactic bar.
The simulation shows that the effect on the angle is larger than the effect on the elongation, producing an observed angle of $\sim$20 degrees, i.e., 5 degrees smaller than the real input one, already at 5$\%$ errors. In our case, the percent error on the distance is indeed very close to 5$\%$ (i.e., an error of $\sim$0.45\,kpc at a distance of 8\,kpc, as shown in Fig.~\ref{fig:err_d} ), which means that our observed angle of 15 degrees might be a few degrees smaller than the real one, closer to 20 degrees. On the other hand, if the same parameter is derived by means of RC stars, with a typical error of 10$\%$ or larger \citep{girardi16}, then the observed angle could easily be half of the real one.
As for the axis ratio, the effect is smaller. In our case, we should have recovered a value identical, within the errors, to the real one, while for RC stars, the bias would be closer to 0.1 if the true axis ratio is 0.7, as in this example.
Of course, a proper simulation needs to be run for each case, iterating on the input parameters until the observed ones are recovered.   

Summarizing, our best estimate for the shape of the structure traced by bulge RRab stars is a spheroid (we may call it a bar) with axis ratios b/a$\sim$0.7 and inclination with respect to the Sun-GC direction of $\sim$20 degrees.  
This structure is less elongated than the one reported by \citet{pietrukowics+15}, having an axis ratio of b/a$\sim$0.49, though not completely axisymmetric as quoted by \citet{dekany+13}.
Taken at face values, the parameters derived here would indicate a bar that is less pronounced than the one traced by means of RC stars. The latter has been reported as having an axis ratio of 0.6 according to \citet{wegg+13}, and of 0.43 and 0.44 according to \citet{cao+13} and \citet{simion+17}, respectively. The inclination angle, on the other hand, is estimated between 25 and 30 degrees, according to different authors \citep[see][for a recent review]{shen+20}. However, we should refrain from a direct comparison with values derived without considering the effect of observational errors discussed in the previous paragraph. In addition, as shown by \citet{zoccali+17} and confirmed by \citet{lim+21} with independent data, the metal poor bulge RC stars trace a component that is much more spheroidal than that traced by metal rich RC stars. Therefore, when tracing the bar using RC stars, it is crucial to separate the two components and de-project their 3D distribution independently.
As discussed by \citet{zoccali+18}, and also evident from the figures in \citet{lim+21}, each of the two components makes up about half of the total number of stars in the bulge, and their relative proportion varies strongly across the bulge area. Therefore, as done so far, blindly tracing a mix of the two populations yields structural parameters that are not appropriate for any of the two.

Following the suggestion by \citet{du+20}, we investigated the possibility that RRL with different metallicities might trace different structures. For our sample, however, we did not find evidences in that sense (see Appendix~B).

\section{Summary and Conclusions}
\label{sec:summary}

We provide a new catalog of 16488 RRab in the bulge region comprised within $-10^\circ$$\lesssim $l$\lesssim $$10^\circ$ and $-2.5^\circ$$\lesssim $b$\lesssim $$2.5^\circ$, based on multi epoch VVV PSF photometry. The first few lines of the catalog are given in Table~\ref{table:cat}, while the complete table is published in electronic form at the CDS.
We selected bona fide RRab, among several million candidate variable stars, by means of an automatic classification based on a Random Forest algorithm. However, because the light curves of our candidate variables, close to the midplane, are noisier than their analog in the training set (OGLE-IV variables as seen by VVV), we accepted face value only the variables with a $>$90$\%$ probability of being RRab. Stars with 50-90$\%$ probability were passed through a few other post-classification filters and finally visually inspected. As a result, our catalog has a high purity in comparison with others in the literature, although it is not complete, especially in the range |b|$<$1$^\circ$. 

For the RRab in our catalog, we derived mean magnitudes in J and \ksnosp, reddening based on their mean colors, photometric metallicities based on their light curves, and distances adopting the PLZ recently provided by \citet{prudil+23}. We verified that the latter is the relation that minimizes the observed trend of distance versus metallicity and period, therefore providing a validation of this PLZ for the present bulge sample. We show that other recent PLZ do not perform equally well with our sample.

Finally, we derive the distribution of the RRab in our new catalog, projected on the Galactic plane. The result is an elongated spheroid with axis ratio b/a\,$\sim$\,0.7 (ellipticity e\,$\sim$\,0.7) and an inclination of $\sim$\,20 degrees with respect to the Sun-GC direction. The above parameters describe a component that is slightly
less elongated (i.e., rounder) than the MW main bar, as characterized by studies based on RC stars, and with a slightly smaller inclination angle. Nonetheless, we emphasized the impact that observational biases have on the shape parameters derived here, and also on those provided in the literature from the density of RC stars.

We conclude that observational errors on the distances can have an important impact on the reconstructed shape of RC stars, which have not been taken into account in previous studies. Also, previous studies based on RC stars have analyzed a mix of two populations (metal rich and metal poor), adopting a unique parametrization for both of them. Because we know that they have a different spatial distribution within the Galaxy, the result is most likely not representative of any of the two.  As for RRL, they are a cleaner sample and provide smaller errors on the distances. Therefore, observational errors have a much lower effect on the derived shape parameters. Unfortunately, their drawback is that their completeness is much lower and most likely uneven across the projected Galactic plane, which introduces different biases that are hard to model and correct. Yet, we argue that this is the RRab catalog offering the best combination of large statistics and high purity in the inner bulge, as we visually inspected a large fraction of their light curves, except for those with the largest S/N ($>$60) and probability of being RRab ($>$90$\%$) according to our classifier. As such, we hope it will allow us to progress in our understanding of the oldest stellar component in the core of our Galaxy.

As a final remark, we note that with the present data we cannot exclude that we are dealing with a mix of two component, one belonging to the metal poor spheroid and another one to the metal rich bar. If that was the case, then it would be natural to derive a structure whose parameters are intermediate between the two, as we seem to find here, and in the literature. We argue, however, that contrary to what happens with bulge RC stars, it is not possible to separate the (putative) two RRL components according to their metallicity, as it shows a single peak distribution
centered at [Fe/H]$\sim$$-$1.35 dex (Fig.~\ref{fig:d_feh}). Perhaps a full 6D orbital analysis will allow a proper distinction, and for this reason it is very important to keep working towards producing clean RRL catalogs, reaching as close as possible to the Galactic midplane, where the density is higher, and for which we could slowly gather accurate 3D kinematics \citep[see, e.g.,][]{olivares+24}.

\begin{acknowledgements}
We thank M\'arcio Catelan and Giuseppe Bono, for many useful discussions during the past few years.

Based on observations taken within the ESO VISTA Public Survey VVV, Program ID 179.B-2002, made public at the ESO Archive and through the Cambridge Astronomical Survey Unit (CASU). \\

This work is funded by ANID, Millenium Science Initiative, ICN12\_009 awarded to the Millennium Institute of Astrophysics  (M.A.S.), by the ANID BASAL Center for Astrophysics and Associated Technologies (CATA) through grant FB210003, and by  FONDECYT Regular grant No. 1230731. C. Q. Z. acknowledges support from the National Agency for Research and Development (ANID), Scholarship Program Doctorado Nacional 2021 – 21211884, ANID. A. R. A. acknowledges support from DICYT through grant 062319RA. E. V. acknowledges the Excellence Cluster ORIGINS Funded by the Deutsche Forschungsgemeinschaft (DFG, German Research Foundation) under Germany’s Excellence Strategy – EXC-2094-390783311. J.O.C. acknowledges support from the National Agency for Research and Development (ANID) Doctorado Nacional grant 2021-21210865, ANID. A. V. N. acknowledges support from the National Agency for Research and Development (ANID), Scholarship Program Doctorado Nacional 2020 – 21201226, ANID.

\end{acknowledgements}


\begin{table*}
\caption{A sample of the RRab catalog released at the CDS together with the present paper.}
\label{table:cat}      
\centering
\tiny
\begin{tabular}{c c c c c c c c c c c}
\hline\hline
ID  &  l  &  b  &  K$_{\rm s}$  &  err K$_{\rm s}$ &  J  & err J &  [Fe/H]  & err [Fe/H] &  E(J$-$K$_{\rm s}$) &  err E(J$-$K$_{\rm s}$) \\   
mag & deg & mag &  mag          &   mag            & mag & mag   &   dex    &  dex       &  mag               &   mag \\
\hline
b299\_102\_76628 &  $-$9.868689 & $-$2.079374 & 15.100  & 0.005  & 16.07  & 0.04  & $-$1.35 & 0.15 & 0.71 & 0.05  \\
b299\_104\_41588 &  $-$9.931120 & $-$2.554848 & 14.860  & 0.003  & 15.74  & 0.02  & $-$0.68 & 0.03 & 0.62 & 0.04  \\
& & & & & & & & & & \\
\hline
& & & & & & & & & & \\
& & & & & & & & & & \\
\hline\hline
Amplitude &  S/N  & Period  & flag  & Distance & err Distance &  X &  Y  &  Z  &    RA       &          DEC    \\        
mag &   --  &  days   &  --   & kpc      & kpc          & kpc & kpc & kpc &  deg  & deg \\
\hline
0.341 & 72.2 &  0.4467 & VVV &  9.72 & 0.47 & 9.57 &  $-$1.67 & $-$0.35 & 262.1421509 & $-$38.378113 \\
0.318 & 90.4 &  0.4673 & VVV &  8.68 & 0.42 & 8.54 &  $-$1.50 & $-$0.39 & 262.6053783 & $-$38.692233 \\
& & & & & & & & & & \\
\hline                  
\end{tabular}
\end{table*}

\bibliographystyle{aa}
\bibliography{rrlyraecatalog}

\begin{thebibliography}{88}
\expandafter\ifx\csname natexlab\endcsname\relax\def\natexlab#1{#1}\fi

\bibitem[{{Ambikasaran} {et~al.}(2015){Ambikasaran}, {Foreman-Mackey},
  {Greengard}, {Hogg}, \& {O'Neil}}]{george_lib}
{Ambikasaran}, S., {Foreman-Mackey}, D., {Greengard}, L., {Hogg}, D.~W., \&
  {O'Neil}, M. 2015, IEEE Transactions on Pattern Analysis and Machine
  Intelligence, 38, 252

\bibitem[{{Anderson} {et~al.}(2006){Anderson}, {Bedin}, {Piotto}, {Yadav}, \&
  {Bellini}}]{anderson06}
{Anderson}, J., {Bedin}, L.~R., {Piotto}, G., {Yadav}, R.~S., \& {Bellini}, A.
  2006, \aap, 454, 1029

\bibitem[{{Bellini} {et~al.}(2014){Bellini}, {Anderson}, {van der Marel},
  {Watkins}, {King}, {Bianchini}, {Chanam{\'e}}, {Chandar}, {Cool}, {Ferraro},
  {Ford}, \& {Massari}}]{bellini14}
{Bellini}, A., {Anderson}, J., {van der Marel}, R.~P., {et~al.} 2014, \apj,
  797, 115

\bibitem[{{Bhardwaj}(2022)}]{bhardwaj+22}
{Bhardwaj}, A. 2022, Universe, 8, 122

\bibitem[{{Bhardwaj} {et~al.}(2023){Bhardwaj}, {Marconi}, {Rejkuba}, {de
  Grijs}, {Singh}, {Braga}, {Kanbur}, {Ngeow}, {Ripepi}, {Bono}, {De Somma}, \&
  {Dall'Ora}}]{bhardwaj+23}
{Bhardwaj}, A., {Marconi}, M., {Rejkuba}, M., {et~al.} 2023, \apjl, 944, L51

\bibitem[{{Bobylev} \& {Bajkova}(2021)}]{bobylev+21}
{Bobylev}, V.~V. \& {Bajkova}, A.~T. 2021, Astronomy Reports, 65, 498

\bibitem[{{Braga} {et~al.}(2019){Braga}, {Stetson}, {Bono}, {Dall'Ora},
  {Ferraro}, {Fiorentino}, {Iannicola}, {Inno}, {Marengo}, {Neeley}, {Beaton},
  {Buonanno}, {Calamida}, {Contreras Ramos}, {Chaboyer}, {Fabrizio},
  {Freedman}, {Gilligan}, {Johnston}, {Lub}, {Madore}, {Magurno}, {Marconi},
  {Marinoni}, {Marrese}, {Mateo}, {Matsunaga}, {Minniti}, {Monson}, {Monelli},
  {Nonino}, {Persson}, {Pietrinferni}, {Sneden}, {Storm}, {Walker}, {Valenti},
  \& {Zoccali}}]{braga18}
{Braga}, V.~F., {Stetson}, P.~B., {Bono}, G., {et~al.} 2019, \aap, 625, A1

\bibitem[{{Breiman}(2001)}]{RFcode}
{Breiman}, L. 2001, Machine Learning, 45, 5

\bibitem[{{Cabral} {et~al.}(2020){Cabral}, {Ramos}, {Gurovich}, \&
  {Granitto}}]{cabral+20}
{Cabral}, J.~B., {Ramos}, F., {Gurovich}, S., \& {Granitto}, P.~M. 2020, \aap,
  642, A58

\bibitem[{{Cao} {et~al.}(2013){Cao}, {Mao}, {Nataf}, {Rattenbury}, \&
  {Gould}}]{cao+13}
{Cao}, L., {Mao}, S., {Nataf}, D., {Rattenbury}, N.~J., \& {Gould}, A. 2013,
  \mnras, 434, 595

\bibitem[{{Carpenter} {et~al.}(2001){Carpenter}, {Hillenbrand}, \&
  {Skrutskie}}]{carpenter01}
{Carpenter}, J.~M., {Hillenbrand}, L.~A., \& {Skrutskie}, M.~F. 2001, \aj, 121,
  3160

\bibitem[{{Catelan} \& {Smith}(2015)}]{catelansmith15}
{Catelan}, M. \& {Smith}, H.~A. 2015, {Pulsating Stars}

\bibitem[{{Cava} {et~al.}(2018){Cava}, {Schaerer}, {Richard},
  {P{\'e}rez-Gonz{\'a}lez}, {Dessauges-Zavadsky}, {Mayer}, \&
  {Tamburello}}]{Cava+18}
{Cava}, A., {Schaerer}, D., {Richard}, J., {et~al.} 2018, Nature Astronomy, 2,
  76

\bibitem[{{Clementini} {et~al.}(2022){Clementini}, {Ripepi}, {Garofalo},
  {Molinaro}, {Muraveva}, {Leccia}, {Rimoldini}, {Holl}, {Jevardat de
  Fombelle}, {Sartoretti}, {Marchal}, {Audard}, {Nienartowicz}, {Andrae},
  {Marconi}, {Szabados}, {Evans}, {Lecoeur-Taibi}, {Mowlavi}, {Musella}, \&
  {Eyer}}]{clementini+22}
{Clementini}, G., {Ripepi}, V., {Garofalo}, A., {et~al.} 2022, arXiv e-prints,
  arXiv:2206.06278

\bibitem[{{Contreras Ramos} {et~al.}(2018){Contreras Ramos}, {Minniti}, {Gran},
  {Zoccali}, {Alonso-Garc{\'\i}a}, {Huijse}, {Navarro}, {Rojas-Arriagada}, \&
  {Valenti}}]{contrerasramos18}
{Contreras Ramos}, R., {Minniti}, D., {Gran}, F., {et~al.} 2018, \apj, 863, 79

\bibitem[{{Contreras Ramos} {et~al.}(2017){Contreras Ramos}, {Zoccali},
  {Rojas}, {Rojas-Arriagada}, {G{\'a}rate}, {Huijse}, {Gran}, {Soto},
  {Valcarce}, {Est{\'e}vez}, \& {Minniti}}]{contrerasramos17}
{Contreras Ramos}, R., {Zoccali}, M., {Rojas}, F., {et~al.} 2017, \aap, 608,
  A140

\bibitem[{{Crestani} {et~al.}(2022){Crestani}, {Fabrizio}, {Braga}, {Sneden},
  {Preston}, {Ferraro}, {Iannicola}, {Bono}, {Alves-Brito}, {Nonino},
  {D'Orazi}, {Inno}, {Monelli}, {Storm}, {Altavilla}, {Chaboyer}, {Dall'Ora},
  {Fiorentino}, {Gilligan}, {Grebel}, {Lala}, {Lemasle}, {Marengo}, {Marinoni},
  {Marrese}, {Martinez-Vazquez}, {Matsunaga}, {Mullen}, {Neeley}, {Prudil}, {da
  Silva}, {Stetson}, {Thevenin}, {Valenti}, {Walker}, \&
  {Zoccali}}]{crestani+21}
{Crestani}, J., {Fabrizio}, M., {Braga}, V.~F., {et~al.} 2022, VizieR Online
  Data Catalog, J/ApJ/908/20

\bibitem[{{Daza-Perilla} {et~al.}(2023){Daza-Perilla}, {Gramajo}, {Lares},
  {Palma}, {Ferreira Lopes}, {Minniti}, \& {Clari{\'a}}}]{daza-perilla23}
{Daza-Perilla}, I.~V., {Gramajo}, L.~V., {Lares}, M., {et~al.} 2023, \mnras,
  520, 828

\bibitem[{{Debattista} {et~al.}(2023){Debattista}, {Liddicott}, {Gonzalez},
  {Beraldo e Silva}, {Amarante}, {Lazar}, {Zoccali}, {Valenti}, {Fisher},
  {Khachaturyants}, {Nidever}, {Quinn}, {Du}, \& {Kassin}}]{debattista+23}
{Debattista}, V.~P., {Liddicott}, D.~J., {Gonzalez}, O.~A., {et~al.} 2023,
  \apj, 946, 118

\bibitem[{{Debattista} {et~al.}(2017){Debattista}, {Ness}, {Gonzalez},
  {Freeman}, {Zoccali}, \& {Minniti}}]{debattista+17}
{Debattista}, V.~P., {Ness}, M., {Gonzalez}, O.~A., {et~al.} 2017, \mnras, 469,
  1587

\bibitem[{{D{\'e}k{\'a}ny} \& {Grebel}(2020)}]{dekany+20}
{D{\'e}k{\'a}ny}, I. \& {Grebel}, E.~K. 2020, \apj, 898, 46

\bibitem[{{D{\'e}k{\'a}ny} \& {Grebel}(2022)}]{dekany+22}
{D{\'e}k{\'a}ny}, I. \& {Grebel}, E.~K. 2022, \apjs, 261, 33

\bibitem[{{D{\'e}k{\'a}ny} {et~al.}(2013){D{\'e}k{\'a}ny}, {Minniti},
  {Catelan}, {Zoccali}, {Saito}, {Hempel}, \& {Gonzalez}}]{dekany+13}
{D{\'e}k{\'a}ny}, I., {Minniti}, D., {Catelan}, M., {et~al.} 2013, \apjl, 776,
  L19

\bibitem[{{Dessauges-Zavadsky} {et~al.}(2017){Dessauges-Zavadsky}, {Schaerer},
  {Cava}, {Mayer}, \& {Tamburello}}]{Dessauges+17}
{Dessauges-Zavadsky}, M., {Schaerer}, D., {Cava}, A., {Mayer}, L., \&
  {Tamburello}, V. 2017, \apjl, 836, L22

\bibitem[{{Di Matteo}(2016)}]{diMatteo+16}
{Di Matteo}, P. 2016, \pasa, 33, e027

\bibitem[{{Du} {et~al.}(2020){Du}, {Mao}, {Athanassoula}, {Shen}, \&
  {Pietrukowicz}}]{du+20}
{Du}, H., {Mao}, S., {Athanassoula}, E., {Shen}, J., \& {Pietrukowicz}, P.
  2020, \mnras, 498, 5629

\bibitem[{{Elorrieta} {et~al.}(2016){Elorrieta}, {Eyheramendy}, {Jord{\'a}n},
  {D{\'e}k{\'a}ny}, {Catelan}, {Angeloni}, {Alonso-Garc{\'\i}a},
  {Contreras-Ramos}, {Gran}, {Hajdu}, {Espinoza}, {Saito}, \&
  {Minniti}}]{elorrieta+16}
{Elorrieta}, F., {Eyheramendy}, S., {Jord{\'a}n}, A., {et~al.} 2016, \aap, 595,
  A82

\bibitem[{{Emerson} {et~al.}(2004){Emerson}, {Irwin}, {Lewis}, {Hodgkin},
  {Evans}, {Bunclark}, {McMahon}, {Hambly}, {Mann}, {Bond}, {Sutorius}, {Read},
  {Williams}, {Lawrence}, \& {Stewart}}]{emerson04}
{Emerson}, J.~P., {Irwin}, M.~J., {Lewis}, J., {et~al.} 2004, in Society of
  Photo-Optical Instrumentation Engineers (SPIE) Conference Series, Vol. 5493,
  Optimizing Scientific Return for Astronomy through Information Technologies,
  ed. P.~J. {Quinn} \& A.~{Bridger}, 401--410

\bibitem[{{Fragkoudi} {et~al.}(2018){Fragkoudi}, {Di Matteo}, {Haywood},
  {Schultheis}, {Khoperskov}, {G{\'o}mez}, \& {Combes}}]{fragkoudi+18}
{Fragkoudi}, F., {Di Matteo}, P., {Haywood}, M., {et~al.} 2018, \aap, 616, A180

\bibitem[{{Fragkoudi} {et~al.}(2020){Fragkoudi}, {Grand}, {Pakmor},
  {Bl{\'a}zquez-Calero}, {Gargiulo}, {Gomez}, {Marinacci}, {Monachesi}, {Ness},
  {Perez}, {Tissera}, \& {White}}]{fragkoudi+20}
{Fragkoudi}, F., {Grand}, R.~J.~J., {Pakmor}, R., {et~al.} 2020, \mnras, 494,
  5936

\bibitem[{{Girardi}(2016)}]{girardi16}
{Girardi}, L. 2016, \araa, 54, 95

\bibitem[{{Gonz{\'a}lez-Fern{\'a}ndez}
  {et~al.}(2018){Gonz{\'a}lez-Fern{\'a}ndez}, {Hodgkin}, {Irwin},
  {Gonz{\'a}lez-Solares}, {Koposov}, {Lewis}, {Emerson}, {Hewett},
  {Yolda{\c{s}}}, \& {Riello}}]{gonzalez-fernandez18}
{Gonz{\'a}lez-Fern{\'a}ndez}, C., {Hodgkin}, S.~T., {Irwin}, M.~J., {et~al.}
  2018, \mnras, 474, 5459

\bibitem[{{Gran} {et~al.}(2016){Gran}, {Minniti}, {Saito}, {Zoccali},
  {Gonzalez}, {Navarrete}, {Catelan}, {Contreras Ramos}, {Elorrieta},
  {Eyheramendy}, \& {Jord{\'a}n}}]{gran+16}
{Gran}, F., {Minniti}, D., {Saito}, R.~K., {et~al.} 2016, \aap, 591, A145

\bibitem[{{GRAVITY Collaboration} {et~al.}(2022){GRAVITY Collaboration},
  {Abuter}, {Aimar}, {Amorim}, {Ball}, {Baub{\"o}ck}, {Berger}, {Bonnet},
  {Bourdarot}, {Brandner}, {Cardoso}, {Cl{\'e}net}, {Dallilar}, {Davies}, {de
  Zeeuw}, {Dexter}, {Drescher}, {Eisenhauer}, {F{\"o}rster Schreiber},
  {Foschi}, {Garcia}, {Gao}, {Gendron}, {Genzel}, {Gillessen}, {Habibi},
  {Haubois}, {Hei{\ss}el}, {Henning}, {Hippler}, {Horrobin}, {Jochum}, {Jocou},
  {Kaufer}, {Kervella}, {Lacour}, {Lapeyr{\`e}re}, {Le Bouquin}, {L{\'e}na},
  {Lutz}, {Ott}, {Paumard}, {Perraut}, {Perrin}, {Pfuhl}, {Rabien},
  {Shangguan}, {Shimizu}, {Scheithauer}, {Stadler}, {Stephens}, {Straub},
  {Straubmeier}, {Sturm}, {Tacconi}, {Tristram}, {Vincent}, {von Fellenberg},
  {Widmann}, {Wieprecht}, {Wiezorrek}, {Woillez}, {Yazici}, \&
  {Young}}]{gravity+22}
{GRAVITY Collaboration}, {Abuter}, R., {Aimar}, N., {et~al.} 2022, \aap, 657,
  L12

\bibitem[{{GRAVITY Collaboration} {et~al.}(2021){GRAVITY Collaboration},
  {Abuter}, {Amorim}, {Baub{\"o}ck}, {Berger}, {Bonnet}, {Brandner},
  {Cl{\'e}net}, {Davies}, {de Zeeuw}, {Dexter}, {Dallilar}, {Drescher},
  {Eckart}, {Eisenhauer}, {F{\"o}rster Schreiber}, {Garcia}, {Gao}, {Gendron},
  {Genzel}, {Gillessen}, {Habibi}, {Haubois}, {Hei{\ss}el}, {Henning},
  {Hippler}, {Horrobin}, {Jim{\'e}nez-Rosales}, {Jochum}, {Jocou}, {Kaufer},
  {Kervella}, {Lacour}, {Lapeyr{\`e}re}, {Le Bouquin}, {L{\'e}na}, {Lutz},
  {Nowak}, {Ott}, {Paumard}, {Perraut}, {Perrin}, {Pfuhl}, {Rabien},
  {Rodr{\'\i}guez-Coira}, {Shangguan}, {Shimizu}, {Scheithauer}, {Stadler},
  {Straub}, {Straubmeier}, {Sturm}, {Tacconi}, {Vincent}, {von Fellenberg},
  {Waisberg}, {Widmann}, {Wieprecht}, {Wiezorrek}, {Woillez}, {Yazici},
  {Young}, \& {Zins}}]{gravity+21}
{GRAVITY Collaboration}, {Abuter}, R., {Amorim}, A., {et~al.} 2021, \aap, 647,
  A59

\bibitem[{{Griv} {et~al.}(2020){Griv}, {Gedalin}, {Pietrukowicz}, {Majaess}, \&
  {Jiang}}]{griv+20}
{Griv}, E., {Gedalin}, M., {Pietrukowicz}, P., {Majaess}, D., \& {Jiang}, I.-G.
  2020, \mnras, 499, 1091

\bibitem[{{Guo} {et~al.}(2015){Guo}, {Ferguson}, {Bell}, {Koo}, {Conselice},
  {Giavalisco}, {Kassin}, {Lu}, {Lucas}, {Mandelker}, {McIntosh}, {Primack},
  {Ravindranath}, {Barro}, {Ceverino}, {Dekel}, {Faber}, {Fang}, {Koekemoer},
  {Noeske}, {Rafelski}, \& {Straughn}}]{Guo+15}
{Guo}, Y., {Ferguson}, H.~C., {Bell}, E.~F., {et~al.} 2015, \apj, 800, 39

\bibitem[{{Guo} {et~al.}(2018){Guo}, {Rafelski}, {Bell}, {Conselice}, {Dekel},
  {Faber}, {Giavalisco}, {Koekemoer}, {Koo}, {Lu}, {Mandelker}, {Primack},
  {Ceverino}, {de Mello}, {Ferguson}, {Hathi}, {Kocevski}, {Lucas},
  {P{\'e}rez-Gonz{\'a}lez}, {Ravindranath}, {Soto}, {Straughn}, \&
  {Wang}}]{Guo+18}
{Guo}, Y., {Rafelski}, M., {Bell}, E.~F., {et~al.} 2018, \apj, 853, 108

\bibitem[{{Hajdu} {et~al.}(2020){Hajdu}, {D{\'e}k{\'a}ny}, {Catelan}, \&
  {Grebel}}]{hajdu+20}
{Hajdu}, G., {D{\'e}k{\'a}ny}, I., {Catelan}, M., \& {Grebel}, E.~K. 2020,
  Experimental Astronomy, 49, 217

\bibitem[{{Hajdu} {et~al.}(2018){Hajdu}, {D{\'e}k{\'a}ny}, {Catelan}, {Grebel},
  \& {Jurcsik}}]{hajdu+18}
{Hajdu}, G., {D{\'e}k{\'a}ny}, I., {Catelan}, M., {Grebel}, E.~K., \&
  {Jurcsik}, J. 2018, \apj, 857, 55

\bibitem[{{Hey} {et~al.}(2023){Hey}, {Huber}, {Shappee}, {Bland-Hawthorn},
  {Tepper-Garc{\'\i}a}, {Sanderson}, {Chakrabarti}, {Saunders}, {Hunt},
  {Bedding}, \& {Tonry}}]{hey+23}
{Hey}, D.~R., {Huber}, D., {Shappee}, B.~J., {et~al.} 2023, \aj, 166, 249

\bibitem[{{Huertas-Company} {et~al.}(2020){Huertas-Company}, {Guo}, {Ginzburg},
  {Lee}, {Mandelker}, {Metter}, {Primack}, {Dekel}, {Ceverino}, {Faber}, {Koo},
  {Koekemoer}, {Snyder}, {Giavalisco}, \& {Zhang}}]{Huertas+20}
{Huertas-Company}, M., {Guo}, Y., {Ginzburg}, O., {et~al.} 2020, \mnras, 499,
  814

\bibitem[{{Johnson} {et~al.}(2022){Johnson}, {Rich}, {Simion}, {Young},
  {Clarkson}, {Pilachowski}, {Michael}, {Marchetti}, {Soto}, {Kunder},
  {Koch-Hansen}, {Katherina Vivas}, {Joyce}, {Shen}, \& {Osmond}}]{johnson+22}
{Johnson}, C.~I., {Rich}, R.~M., {Simion}, I.~T., {et~al.} 2022, \mnras, 515,
  1469

\bibitem[{{Kim} {et~al.}(2011){Kim}, {Protopapas}, {Byun}, {Alcock}, {Khardon},
  \& {Trichas}}]{kim11}
{Kim}, D.-W., {Protopapas}, P., {Byun}, Y.-I., {et~al.} 2011, \apj, 735, 68

\bibitem[{{Kinemuchi} {et~al.}(2006){Kinemuchi}, {Smith}, {Wo{\'z}niak},
  {McKay}, \& {ROTSE Collaboration}}]{kinemuchi06}
{Kinemuchi}, K., {Smith}, H.~A., {Wo{\'z}niak}, P.~R., {McKay}, T.~A., \&
  {ROTSE Collaboration}. 2006, \aj, 132, 1202

\bibitem[{{Kormendy} \& {Kennicutt}(2004)}]{kk+04}
{Kormendy}, J. \& {Kennicutt}, Robert~C., J. 2004, \araa, 42, 603

\bibitem[{{Kunder} {et~al.}(2016){Kunder}, {Rich}, {Koch}, {Storm}, {Nataf},
  {De Propris}, {Walker}, {Bono}, {Johnson}, {Shen}, \& {Li}}]{kunder+16}
{Kunder}, A., {Rich}, R.~M., {Koch}, A., {et~al.} 2016, \apjl, 821, L25

\bibitem[{{Leung} {et~al.}(2023){Leung}, {Bovy}, {Mackereth}, {Hunt}, {Lane},
  \& {Wilson}}]{leung+23}
{Leung}, H.~W., {Bovy}, J., {Mackereth}, J.~T., {et~al.} 2023, \mnras, 519, 948

\bibitem[{{Lim} {et~al.}(2021){Lim}, {Koch-Hansen}, {Chung}, {Johnson},
  {Kunder}, {Simion}, {Rich}, {Clarkson}, {Pilachowski}, {Michael}, {Vivas}, \&
  {Young}}]{lim+21}
{Lim}, D., {Koch-Hansen}, A.~J., {Chung}, C., {et~al.} 2021, \aap, 647, A34

\bibitem[{{Majaess} {et~al.}(2018){Majaess}, {D{\'e}k{\'a}ny}, {Hajdu},
  {Minniti}, {Turner}, \& {Gieren}}]{majaess+18}
{Majaess}, D., {D{\'e}k{\'a}ny}, I., {Hajdu}, G., {et~al.} 2018, \apss, 363,
  127

\bibitem[{{Mart{\'\i}nez-V{\'a}zquez}
  {et~al.}(2017){Mart{\'\i}nez-V{\'a}zquez}, {Monelli}, {Bernard}, {Gallart},
  {Stetson}, {Skillman}, {Bono}, {Cassisi}, {Fiorentino}, {McQuinn}, {Cole},
  {McConnachie}, {Martin}, {Dolphin}, {Boylan-Kolchin}, {Aparicio}, {Hidalgo},
  \& {Weisz}}]{martinez-vazquez+17}
{Mart{\'\i}nez-V{\'a}zquez}, C.~E., {Monelli}, M., {Bernard}, E.~J., {et~al.}
  2017, \apj, 850, 137

\bibitem[{{Medina} {et~al.}(2018){Medina}, {Borissova}, {Bayo}, {Kurtev},
  {Navarro Molina}, {Kuhn}, {Kumar}, {Lucas}, {Catelan}, {Minniti}, \&
  {Smith}}]{medina+18}
{Medina}, N., {Borissova}, J., {Bayo}, A., {et~al.} 2018, \apj, 864, 11

\bibitem[{{Minniti} {et~al.}(2010){Minniti}, {Lucas}, {Emerson}, {Saito},
  {Hempel}, {Pietrukowicz}, {Ahumada}, {Alonso}, {Alonso-Garcia}, {Arias},
  {Bandyopadhyay}, {Barb{\'a}}, {Barbuy}, {Bedin}, {Bica}, {Borissova},
  {Bronfman}, {Carraro}, {Catelan}, {Clari{\'a}}, {Cross}, {de Grijs},
  {D{\'e}k{\'a}ny}, {Drew}, {Fari{\~n}a}, {Feinstein}, {Fern{\'a}ndez
  Laj{\'u}s}, {Gamen}, {Geisler}, {Gieren}, {Goldman}, {Gonzalez}, {Gunthardt},
  {Gurovich}, {Hambly}, {Irwin}, {Ivanov}, {Jord{\'a}n}, {Kerins}, {Kinemuchi},
  {Kurtev}, {L{\'o}pez-Corredoira}, {Maccarone}, {Masetti}, {Merlo},
  {Messineo}, {Mirabel}, {Monaco}, {Morelli}, {Padilla}, {Palma}, {Parisi},
  {Pignata}, {Rejkuba}, {Roman-Lopes}, {Sale}, {Schreiber}, {Schr{\"o}der},
  {Smith}, {}, {Soto}, {Tamura}, {Tappert}, {Thompson}, {Toledo}, {Zoccali}, \&
  {Pietrzynski}}]{minniti10}
{Minniti}, D., {Lucas}, P.~W., {Emerson}, J.~P., {et~al.} 2010, \na, 15, 433

\bibitem[{{Minniti} {et~al.}(2020){Minniti}, {Sbordone}, {Rojas-Arriagada},
  {Zoccali}, {Contreras Ramos}, {Minniti}, {Marconi}, {Braga}, {Catelan},
  {Duffau}, {Gieren}, \& {Valcarce}}]{minniti+20}
{Minniti}, J.~H., {Sbordone}, L., {Rojas-Arriagada}, A., {et~al.} 2020, \aap,
  640, A92

\bibitem[{{Molnar} {et~al.}(2022){Molnar}, {Sanders}, {Smith}, {Belokurov},
  {Lucas}, \& {Minniti}}]{molnar+22}
{Molnar}, T.~A., {Sanders}, J.~L., {Smith}, L.~C., {et~al.} 2022, \mnras, 509,
  2566

\bibitem[{{Muraveva} {et~al.}(2018){Muraveva}, {Delgado}, {Clementini},
  {Sarro}, \& {Garofalo}}]{muraveva18}
{Muraveva}, T., {Delgado}, H.~E., {Clementini}, G., {Sarro}, L.~M., \&
  {Garofalo}, A. 2018, \mnras, 481, 1195

\bibitem[{{Navarrete} {et~al.}(2015){Navarrete}, {Contreras Ramos}, {Catelan},
  {Clement}, {Gran}, {Alonso-Garc{\'\i}a}, {Angeloni}, {Hempel},
  {D{\'e}k{\'a}ny}, \& {Minniti}}]{navarrete15}
{Navarrete}, C., {Contreras Ramos}, R., {Catelan}, M., {et~al.} 2015, \aap,
  577, A99

\bibitem[{{Neeley} {et~al.}(2019){Neeley}, {Marengo}, {Freedman}, {Madore},
  {Beaton}, {Hatt}, {Hoyt}, {Monson}, {Rich}, {Sarajedini}, {Seibert}, \&
  {Scowcroft}}]{neely+19}
{Neeley}, J.~R., {Marengo}, M., {Freedman}, W.~L., {et~al.} 2019, \mnras, 490,
  4254

\bibitem[{{Ness} {et~al.}(2012){Ness}, {Freeman}, {Athanassoula},
  {Wylie-De-Boer}, {Bland-Hawthorn}, {Lewis}, {Yong}, {Asplund}, {Lane},
  {Kiss}, \& {Ibata}}]{ness+12}
{Ness}, M., {Freeman}, K., {Athanassoula}, E., {et~al.} 2012, \apj, 756, 22

\bibitem[{{Nikzat} {et~al.}(2022){Nikzat}, {Ferreira Lopes}, {Catelan},
  {Contreras Ramos}, {Zoccali}, {Rojas-Arriagada}, {Braga}, {Minniti},
  {Borissova}, \& {Becker}}]{nikzat22}
{Nikzat}, F., {Ferreira Lopes}, C.~E., {Catelan}, M., {et~al.} 2022, \aap, 660,
  A35

\bibitem[{{Nun} {et~al.}(2015){Nun}, {Protopapas}, {Sim}, {Zhu}, {Dave},
  {Castro}, \& {Pichara}}]{nun+15}
{Nun}, I., {Protopapas}, P., {Sim}, B., {et~al.} 2015, arXiv e-prints,
  arXiv:1506.00010

\bibitem[{{Olivares Carvajal} {et~al.}(2024){Olivares Carvajal}, {Zoccali}, {De
  Leo}, {Contreras Ramos}, {Quezada}, {Rojas-Arriagada}, {Valenti},
  {Albarrac{\'\i}n}, \& {Valenzuela Navarro}}]{olivares+24}
{Olivares Carvajal}, J., {Zoccali}, M., {De Leo}, M., {et~al.} 2024, arXiv
  e-prints, arXiv:2405.08990

\bibitem[{Pedregosa {et~al.}(2011)Pedregosa, Varoquaux, Gramfort, Michel,
  Thirion, Grisel, Blondel, Prettenhofer, Weiss, Dubourg, Vanderplas, Passos,
  Cournapeau, Brucher, Perrot, \& Duchesnay}]{scikit-learn}
Pedregosa, F., Varoquaux, G., Gramfort, A., {et~al.} 2011, Journal of Machine
  Learning Research, 12, 2825

\bibitem[{{Pietrukowicz} {et~al.}(2015){Pietrukowicz}, {Koz{\l}owski},
  {Skowron}, {Soszy{\'n}ski}, {Udalski}, {Poleski}, {Wyrzykowski},
  {Szyma{\'n}ski}, {Pietrzy{\'n}ski}, {Ulaczyk}, {Mr{\'o}z}, {Skowron}, \&
  {Kubiak}}]{pietrukowics+15}
{Pietrukowicz}, P., {Koz{\l}owski}, S., {Skowron}, J., {et~al.} 2015, \apj,
  811, 113

\bibitem[{{Pietrukowicz} {et~al.}(2012){Pietrukowicz}, {Udalski},
  {Soszy{\'n}ski}, {Nataf}, {Wyrzykowski}, {Poleski}, {Koz{\l}owski},
  {Szyma{\'n}ski}, {Kubiak}, {Pietrzy{\'n}ski}, \& {Ulaczyk}}]{pietrukowicz+12}
{Pietrukowicz}, P., {Udalski}, A., {Soszy{\'n}ski}, I., {et~al.} 2012, \apj,
  750, 169

\bibitem[{{Pietrukowicz} {et~al.}(2020){Pietrukowicz}, {Udalski},
  {Soszy{\'n}ski}, {Skowron}, {Wrona}, {Szyma{\'n}ski}, {Poleski}, {Ulaczyk},
  {Koz{\l}owski}, {Skowron}, {Mr{\'o}z}, {Rybicki}, {Iwanek}, \&
  {Gromadzki}}]{pietrukowicz+20}
{Pietrukowicz}, P., {Udalski}, A., {Soszy{\'n}ski}, I., {et~al.} 2020, \actaa,
  70, 121

\bibitem[{{Prudil} {et~al.}(2023){Prudil}, {Kunder}, {Dekany}, \&
  {Koch-Hansen}}]{prudil+23}
{Prudil}, Z., {Kunder}, A., {Dekany}, I., \& {Koch-Hansen}, A.~J. 2023, arXiv
  e-prints, arXiv:2310.19438

\bibitem[{{Queiroz} {et~al.}(2021){Queiroz}, {Chiappini}, {Perez-Villegas},
  {Khalatyan}, {Anders}, {Barbuy}, {Santiago}, {Steinmetz}, {Cunha},
  {Schultheis}, {Majewski}, {Minchev}, {Minniti}, {Beaton}, {Cohen}, {da
  Costa}, {Fern{\'a}ndez-Trincado}, {Garcia-Hern{\'a}ndez}, {Geisler},
  {Hasselquist}, {Lane}, {Nitschelm}, {Rojas-Arriagada}, {Roman-Lopes},
  {Smith}, \& {Zasowski}}]{queiroz+21}
{Queiroz}, A.~B.~A., {Chiappini}, C., {Perez-Villegas}, A., {et~al.} 2021,
  \aap, 656, A156

\bibitem[{{Richards} {et~al.}(2011){Richards}, {Starr}, {Butler}, {Bloom},
  {Brewer}, {Crellin-Quick}, {Higgins}, {Kennedy}, \& {Rischard}}]{richards+11}
{Richards}, J.~W., {Starr}, D.~L., {Butler}, N.~R., {et~al.} 2011, \apj, 733,
  10

\bibitem[{{Rojas-Arriagada} {et~al.}(2014){Rojas-Arriagada}, {Recio-Blanco},
  {Hill}, {de Laverny}, {Schultheis}, {Babusiaux}, {Zoccali}, {Minniti},
  {Gonzalez}, {Feltzing}, {Gilmore}, {Randich}, {Vallenari}, {Alfaro},
  {Bensby}, {Bragaglia}, {Flaccomio}, {Lanzafame}, {Pancino}, {Smiljanic},
  {Bergemann}, {Costado}, {Damiani}, {Hourihane}, {Jofr{\'e}}, {Lardo},
  {Magrini}, {Maiorca}, {Morbidelli}, {Sbordone}, {Worley}, {Zaggia}, \&
  {Wyse}}]{rojas-arriagada+14}
{Rojas-Arriagada}, A., {Recio-Blanco}, A., {Hill}, V., {et~al.} 2014, \aap,
  569, A103

\bibitem[{{Schwarzenberg-Czerny}(1989)}]{schwarzenberg-czerny89}
{Schwarzenberg-Czerny}, A. 1989, \mnras, 241, 153

\bibitem[{{Semczuk} {et~al.}(2022){Semczuk}, {Dehnen}, {Sch{\"o}nrich}, \&
  {Athanassoula}}]{semczuk+21}
{Semczuk}, M., {Dehnen}, W., {Sch{\"o}nrich}, R., \& {Athanassoula}, E. 2022,
  \mnras, 509, 4532

\bibitem[{{Shen} \& {Zheng}(2020)}]{shen+20}
{Shen}, J. \& {Zheng}, X.-W. 2020, Research in Astronomy and Astrophysics, 20,
  159

\bibitem[{{Simion} {et~al.}(2017){Simion}, {Belokurov}, {Irwin}, {Koposov},
  {Gonzalez-Fernandez}, {Robin}, {Shen}, \& {Li}}]{simion+17}
{Simion}, I.~T., {Belokurov}, V., {Irwin}, M., {et~al.} 2017, \mnras, 471, 4323

\bibitem[{{Smith}(1995)}]{smith95}
{Smith}, H.~A. 1995, Cambridge Astrophysics Series, 27

\bibitem[{{Soszy{\'n}ski} {et~al.}(2011){Soszy{\'n}ski}, {Dziembowski},
  {Udalski}, {Poleski}, {Szyma{\'n}ski}, {Kubiak}, {Pietrzy{\'n}ski},
  {Wyrzykowski}, {Ulaczyk}, {Koz{\l}owski}, \& {Pietrukowicz}}]{ogleIII}
{Soszy{\'n}ski}, I., {Dziembowski}, W.~A., {Udalski}, A., {et~al.} 2011,
  \actaa, 61, 1

\bibitem[{{Soszy{\'n}ski} {et~al.}(2019){Soszy{\'n}ski}, {Udalski}, {Wrona},
  {Szyma{\'n}ski}, {Pietrukowicz}, {Skowron}, {Skowron}, {Poleski},
  {Koz{\l}owski}, {Mr{\'o}z}, {Ulaczyk}, {Rybicki}, {Iwanek}, \&
  {Gromadzki}}]{soszynski+19}
{Soszy{\'n}ski}, I., {Udalski}, A., {Wrona}, M., {et~al.} 2019, \actaa, 69, 321

\bibitem[{{Stetson}(1987)}]{stetson87}
{Stetson}, P.~B. 1987, \pasp, 99, 191

\bibitem[{{Stringer} {et~al.}(2021){Stringer}, {Drlica-Wagner}, {Macri},
  {Mart{\'\i}nez-V{\'a}zquez}, {Vivas}, {Ferguson}, {Pace}, {Walker},
  {Neilsen}, {Tavangar}, {Wester}, {Abbott}, {Aguena}, {Allam}, {Bacon},
  {Bechtol}, {Bertin}, {Brooks}, {Burke}, {Carnero Rosell}, {Carrasco Kind},
  {Carretero}, {Costanzi}, {Crocce}, {da Costa}, {Pereira}, {De Vicente},
  {Desai}, {Diehl}, {Doel}, {Ferrero}, {Garc{\'\i}a-Bellido}, {Gaztanaga},
  {Gerdes}, {Gruen}, {Gruendl}, {Gschwend}, {Gutierrez}, {Hinton}, {Hollowood},
  {Honscheid}, {Hoyle}, {James}, {Kuehn}, {Kuropatkin}, {Li}, {Maia},
  {Marshall}, {Menanteau}, {Miquel}, {Morgan}, {Ogando}, {Palmese},
  {Paz-Chinch{\'o}n}, {Plazas}, {Roodman}, {Sanchez}, {Schubnell}, {Serrano},
  {Sevilla-Noarbe}, {Smith}, {Soares-Santos}, {Suchyta}, {Tarle}, {Thomas},
  {To}, {Varga}, {Wilkinson}, {Zhang}, \& {DES Collaboration}}]{stringer+21}
{Stringer}, K.~M., {Drlica-Wagner}, A., {Macri}, L., {et~al.} 2021, \apj, 911,
  109

\bibitem[{{Stringer} {et~al.}(2019){Stringer}, {Long}, {Macri}, {Marshall},
  {Drlica-Wagner}, {Mart{\'\i}nez-V{\'a}zquez}, {Vivas}, {Bechtol},
  {Morganson}, {Carrasco Kind}, {Pace}, {Walker}, {Nielsen}, {Li}, {Rykoff},
  {Burke}, {Carnero Rosell}, {Neilsen}, {Ferguson}, {Cantu}, {Myron},
  {Strigari}, {Farahi}, {Paz-Chinch{\'o}n}, {Tucker}, {Lin}, {Hatt}, {Maner},
  {Plybon}, {Riley}, {Nadler}, {Abbott}, {Allam}, {Annis}, {Bertin}, {Brooks},
  {Buckley-Geer}, {Carretero}, {Cunha}, {D'Andrea}, {da Costa}, {De Vicente},
  {Desai}, {Doel}, {Eifler}, {Flaugher}, {Frieman}, {Garc{\'\i}a-Bellido},
  {Gaztanaga}, {Gruen}, {Gschwend}, {Gutierrez}, {Hartley}, {Hollowood},
  {Hoyle}, {James}, {Kuehn}, {Kuropatkin}, {Melchior}, {Miquel}, {Ogando},
  {Plazas}, {Sanchez}, {Santiago}, {Scarpine}, {Schubnell}, {Serrano},
  {Sevilla-Noarbe}, {Smith}, {Smith}, {Soares-Santos}, {Sobreira}, {Suchyta},
  {Swanson}, {Tarle}, {Thomas}, {Vikram}, {Yanny}, \& {DES
  Collaboration}}]{stringer+19}
{Stringer}, K.~M., {Long}, J.~P., {Macri}, L.~M., {et~al.} 2019, \aj, 158, 16

\bibitem[{{Surot} {et~al.}(2020){Surot}, {Valenti}, {Gonzalez}, {Zoccali},
  {S{\"o}kmen}, {Hidalgo}, \& {Minniti}}]{surot20}
{Surot}, F., {Valenti}, E., {Gonzalez}, O.~A., {et~al.} 2020, \aap, 644, A140

\bibitem[{{Udalski} {et~al.}(2015){Udalski}, {Szyma{\'n}ski}, \&
  {Szyma{\'n}ski}}]{ogleIV}
{Udalski}, A., {Szyma{\'n}ski}, M.~K., \& {Szyma{\'n}ski}, G. 2015, \actaa, 65,
  1

\bibitem[{{Vislosky} {et~al.}(2024){Vislosky}, {Minchev}, {Khoperskov},
  {Martig}, {Buck}, {Hilmi}, {Ratcliffe}, {Bland-Hawthorn}, {Quillen},
  {Steinmetz}, \& {de Jong}}]{vislosky+24}
{Vislosky}, E., {Minchev}, I., {Khoperskov}, S., {et~al.} 2024, \mnras, 528,
  3576

\bibitem[{{Walker}(1989)}]{walker+89}
{Walker}, A.~R. 1989, \pasp, 101, 570

\bibitem[{{Wegg} \& {Gerhard}(2013)}]{wegg+13}
{Wegg}, C. \& {Gerhard}, O. 2013, \mnras, 435, 1874

\bibitem[{{Zgirski} {et~al.}(2023){Zgirski}, {Pietrzy{\'n}ski}, {G{\'o}rski},
  {Gieren}, {Wielg{\'o}rski}, {Karczmarek}, {Hajdu}, {Lewis}, {Chini},
  {Graczyk}, {Ka{\l}uszy{\'n}ski}, {Narloch}, {Pilecki}, {Rojas Garc{\'\i}a},
  {Suchomska}, \& {Taormina}}]{bartolomeo+23}
{Zgirski}, B., {Pietrzy{\'n}ski}, G., {G{\'o}rski}, M., {et~al.} 2023, arXiv
  e-prints, arXiv:2305.09414

\bibitem[{{Zoccali} {et~al.}(2018){Zoccali}, {Valenti}, \&
  {Gonzalez}}]{zoccali+18}
{Zoccali}, M., {Valenti}, E., \& {Gonzalez}, O.~A. 2018, \aap, 618, A147

\bibitem[{{Zoccali} {et~al.}(2017){Zoccali}, {Vasquez}, {Gonzalez}, {Valenti},
  {Rojas-Arriagada}, {Minniti}, {Rejkuba}, {Minniti}, {McWilliam}, {Babusiaux},
  {Hill}, \& {Renzini}}]{zoccali+17}
{Zoccali}, M., {Vasquez}, S., {Gonzalez}, O.~A., {et~al.} 2017, \aap, 599, A12

\end{thebibliography}


\begin{appendix} 
\section{Post-classification selection}

We illustrate here the criteria used in the post-RF selection. Let us recall that this further selection was aimed at discarding
candidates that, though having probability of being RRab between 50$\%$ and 90$\%$, differ from the golden sample with Prob>90$\%$ and S/N$>$60 in some new parameters. After this selection, the remaining candidates were all visually inspected.

\begin{figure}[hb]
\includegraphics[width=\hsize]{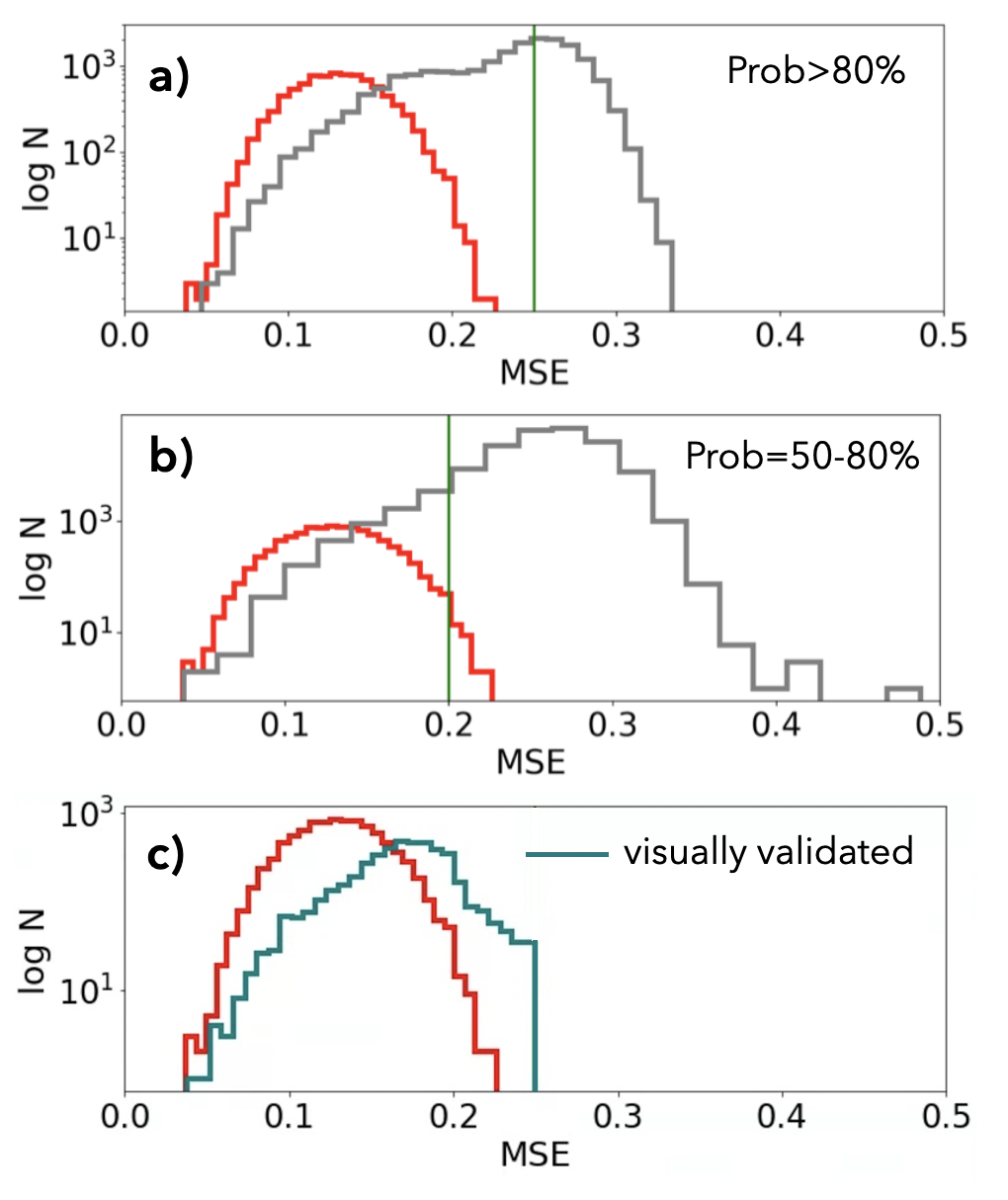}
\caption{We show here that the selection of candidates with Prob$>$80$\%$ and MSE$<$0.25 (top) or Prob=50-80$\%$ and MSE$<$0.2 (middle) allows us to isolate a sub-sample occupying the same parameter space of the golden sample with Prob$>$90$\%$ and S/N$>$60, here shown in red, that we use as a reference.  The bottom panel shows the final sample resulting from the visual inspection (teal) compared with the reference sample (red).}
\label{Fig:post-RF}
\end{figure}

\begin{figure}[ht]
\includegraphics[width=\hsize]{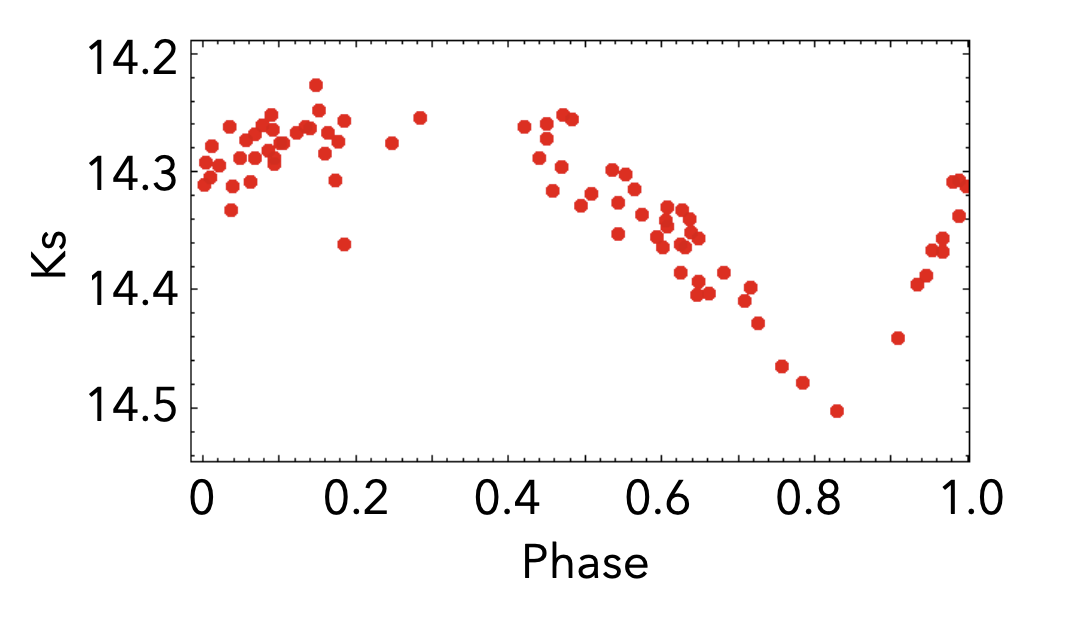}
\caption{Example of a phased light curve that was discarded, for having Nbin=4. That is, if we bin the light curve with 20 bins along the Phase, 4 of them would be empty.}
\label{Fig:Nbin_4}
\end{figure}

\section{The structure traced by RRL with different metallicity}

\begin{figure*}[hb]
\includegraphics[width=\hsize]{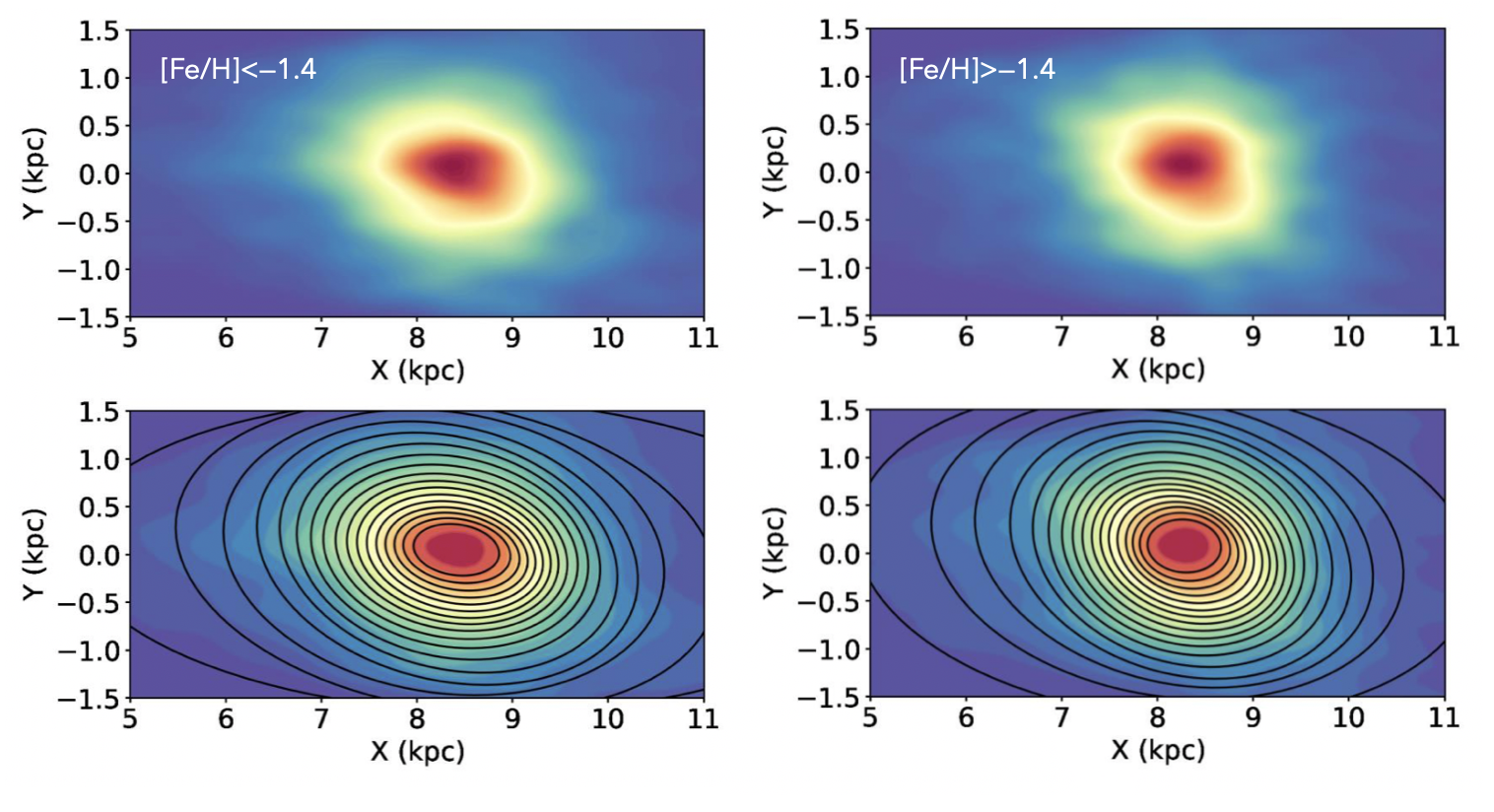}
\caption{Same as Fig.~\ref{fig:KDE}, for the metal poor RRL (left) and the metal rich ones (right).
}
\label{Fig:KDE_feh}
\end{figure*}

Different authors suggested that metal poor RRL trace a more axisymmetric structure than metal rich one, with the latter following more closely the structure of the main Galactic bar \citep{du+20}. In Fig.~\ref{Fig:KDE_feh} shows our attempt at reproducing their results, by dividing the RRab distribution at [Fe/H]=$-1.4$, close to the peak of their metallicity distribution. No significant difference is seen between the two distribution. While the innermost contours might indicate a rounder shape for the metal rich RRL, opposite to previous results, a closer look indicate that this is just due to lower statistics making the axis ratio and inclination angle more stochastic. Indeed, both distributions show a larger variation of the parameters fitted to different contours. 

Moving the metallicity cut does not significantly change the result. Nonetheless, we should note that if there were indeed two components with different metallicities, it is hard to believe that the correct separation between the two would coincide with the peak of the distribution. We should rather look for a dip in the distribution. Such feature has not been seen so far.  Yet, if the separation is closer to one of the two edges of the metallicity distribution, then one of the two component would have too low statistics to be used as a robust tracer.

\end{appendix}

\end{document}